\documentclass[twocolumn,apj,appendixfloats,numberedappendix]{emulateapj}
\usepackage{hyperref}
\usepackage{amsfonts,amsmath,amssymb,amsthm}
\usepackage{longtable}
\usepackage{graphicx}
\usepackage{tabularx}
\usepackage{color}
\usepackage[utf8]{inputenc}
\usepackage[T1]{fontenc}
\usepackage{lmodern}
\usepackage[right]{rotating}
\usepackage{fancyref}
\usepackage{tabularx}
\usepackage{textcomp}

\usepackage[sort&compress]{natbib}
\defcitealias{Franci_al_2015a}{Letter~1}

\newcolumntype{L}[1]{>{\raggedright\arraybackslash}p{#1}}
\newcolumntype{C}[1]{>{\centering\arraybackslash}p{#1}}
\newcolumntype{R}[1]{>{\raggedleft\arraybackslash}p{#1}}
\newcolumntype{Y}{>{\centering\arraybackslash}X}

\newcommand{\vect}[1]{{\boldsymbol{#1}}}

\newcommand{\bnabla}{\boldsymbol{\nabla}}

\shorttitle{Kinetic plasma turbulence at proton scales}
\shortauthors{FRANCI ET AL.}

\begin{document}

\title{High-resolution hybrid simulations of kinetic plasma turbulence at proton scales}

\author{Luca Franci\altaffilmark{1,2}, Simone Landi\altaffilmark{1,3}, Lorenzo Matteini\altaffilmark{4,1}, Andrea Verdini\altaffilmark{1,5}, Petr Hellinger\altaffilmark{6}}
\altaffiltext{1}{Dipartimento di Fisica e Astronomia, Università degli Studi di Firenze, Largo E. Fermi 2, I-50125 Firenze, Italy}
\altaffiltext{2}{INFN - Sezione di Firenze, Via G. Sansone 1, I-50019 Sesto F.no (Firenze), Italy}
\altaffiltext{3}{INAF - Osservatorio Astrofisico di Arcetri, Largo E. Fermi 5, I-50125 Firenze, Italy}
\altaffiltext{4}{Space and Atmospheric Physics Group, Imperial College London, London SW7 2AZ, UK}
\altaffiltext{5}{Solar–Terrestrial Center of Excellence, Royal Observatory of Belgium, Brussels, Belgium}
\altaffiltext{6}{Astronomical Institute, AS CR, Bocni II/1401, CZ-14100 Prague, Czech Republic}
\date{\today}

\begin{abstract}
We investigate properties of plasma turbulence from
magneto-hydrodynamic (MHD) to sub-ion scales by means of
two-dimensional, high-resolution hybrid particle-in-cell
simulations. We impose an initial ambient magnetic field,
perpendicular to the simulation box, and we add a spectrum of
large-scale magnetic and kinetic fluctuations, with energy
equipartition and vanishing correlation.  Once the turbulence is fully
developed, we observe a MHD inertial range, where the spectra of the
perpendicular magnetic field and the perpendicular proton bulk
velocity fluctuations exhibit power-law scaling with spectral indices
of $-5/3$ and $-3/2$, respectively.  This behavior is extended over a
full decade in wavevectors and is very stable in time.  A transition
is observed around proton scales. At sub-ion scales, both spectra
steepen, with the former still following a power law with a spectral
index of $\sim-3$.  A $-2.8$ slope is observed in the density and
parallel magnetic fluctuations, highlighting the presence of
compressive effects at kinetic scales.  The spectrum of the
perpendicular electric fluctuations follows that of the proton bulk
velocity at MHD scales, and flattens at small scales.  All these
features, which we carefully tested against variations of many
parameters, are in good agreement with solar wind observations. The
turbulent cascade leads to on overall proton energization with similar
heating rates in the parallel and perpendicular directions.  While the
parallel proton heating is found to be independent on the resistivity,
the number of particles per cell and the resolution employed, the
perpendicular proton temperature strongly depends on these parameters.
\end{abstract}
\LTcapwidth=\columnwidth 


\maketitle

\section{Introduction}
\label{sec:introduction}

Turbulence is an ubiquitous phenomenon in space and astrophysical
plasmas. Although it is generally driven by violent events or
instabilities at large scales, a further cascade is responsible for
transferring energy via nonlinear coupling from the large injection
scale to much smaller scales, through the ion and the electron
characteristic regimes, where they are eventually dissipated. 
In-situ measurements in the solar wind represent a unique opportunity
to study those processes, since they provide observations in a huge
range of scales (see for example reviews by \citet{Tu_Marsch_1995,
  Matthaeus_Velli_2011, Alexandrova_al_2013, Bruno_Carbone_2013}).
The estimated turbulent energy cascade rate
\citep[e.g.,][]{MacBride_al_2008, Cranmer_al_2009, Hellinger_al_2011,
  Hellinger_al_2013} is comparable to the proton heating needed to
explain the non adiabatic evolution of the solar wind plasma during
its expansion \citep[e.g.,][]{Marsch_al_2004,Matteini_al_2007}.  This
suggests that turbulence plays an active role in transferring energy
from electromagnetic fields to particles and heats the
solar wind plasma.  However, the processes that ultimately lead to
heating in a collisionless turbulent medium are still unknown.

Supporting evidence of a turbulent cascade is provided by the observed
energy spectra, which exhibit a power-law behavior over a large range
of scales, spanning nearly four decades in frequency.  The spectral
index of magnetic and kinetic spectra varies with the temperature of 
the solar wind streams \citep{Grappin_al_1990,Grappin_al_1991}. The latter 
is in turn correlated with the stream speed and, although to a smaller extent,
with the degree of Alfvenicity, i.e., the correlation between kinetic
and magnetic fluctuations \citep{Podesta_Borovsky_2010,
  Chen_al_2013b}.  
However, on average, at fluid-like scales a typical Kolmogorov
power law with spectral index $-5/3$ is usually observed for magnetic
fluctuations, while kinetic energy spectra show a Iroshnikov-Kraichnan
$-3/2$ scaling \citep{Podesta_al_2006a, Podesta_al_2007,
  Tessein_al_2009, Salem_al_2009, Chen_al_2011a, Wicks_al_2011}.
In particular, such scaling is found to be typical
of regimes with balanced turbulence, i.e., zero cross-helicity
\citep{Podesta_Borovsky_2010}.
In the same range of scales, a certain amount of residual energy,
i.e., an excess of magnetic to kinetic energy, is typically observed,
following a well defined power-law scaling with an index of $-2$
\citep{Chen_al_2013b}. The electric field spectrum is observed to
follow the velocity spectrum, when measured in the solar wind frame
\citep{Chen_al_2011b}, while density fluctuations exhibit a
Kolmogorov-like cascade.

In the vicinity of the ion inertial length scale, a break in the
magnetic field power spectrum is observed
\citep{Beinroth_Neubauer_1981, Goldstein_al_1994,Leamon_al_1998,
  Leamon_al_1999}.  Early observations of the spectrum of magnetic
fluctuations in a restricted region above the break found a power-law
scaling with a variable spectral index, ranging from $-2$ to $-4$
\citep[e.g.,][]{Leamon_al_1998, Leamon_al_1999, Bale_al_2005,
  Smith_al_2006b, Alexandrova_al_2008a, Alexandrova_al_2008b,
  Kiyani_al_2009, Sahraoui_al_2009, Chen_al_2010b,
  Salem_al_2012}. However, more recently, observations extended to
smaller scales suggest a general convergence of the spectra towards a
spectral index of $-2.8$ \citep{Kiyani_al_2009, Alexandrova_al_2009,
  Sahraoui_al_2010}, or towards a power-law scaling of $-8/3$,
exponentially damped at sub-electron scales
\citep{Alexandrova_al_2012}. Magnetic fluctuations at sub proton
scales are also characterized by a reduction of the magnetic variance
anisotropy \citep{Podesta_TenBarge_2012}, and by an increase of the
magnetic compressibility \citep{Alexandrova_al_2008b, Salem_al_2012,
  Kiyani_al_2013}, suggesting a change in the nonlinear interactions
ruling the cascade. This is partially confirmed by the measured
increase of the intermittency at ion scales
\citep{Alexandrova_al_2008b, Kiyani_al_2009, Kiyani_al_2013,
  Wu_al_2013, Chen_al_2014a}, although a clear behavior of the
flatness at smaller, sub-ion, scales has not been identified yet.

There are observational indications that, below the ion inertial
length scale, the electric field spectrum decouples from the velocity
field and flattens \citep{Bale_al_2005, Salem_al_2012} but, due to the
high noise level, present data do not allow to determine the existence
of a power-law scaling at sub-ion scales. Density fluctuations show a
plateau just before the ion scales, while they follow a power law between
the ion and the electron scales, with the same spectral index as the one of
the magnetic field spectrum \citep{Chen_al_2012, Chen_al_2013}.

Properties of turbulence have been extensively analyzed by means of
direct numerical simulations (DNS), employing many different methods
and models.  Although several features of the solar wind turbulence can be
partially recovered, we are still far from a comprehensive picture.
At large fluid-like scales, DNS of incompressible MHD and reduced MHD
(RMHD) return a spectral index for the total energy close to $-2$,
$-5/3$, or $-3/2$ \citep[e.g.,][]{Maron_Goldreich_2001,
  Muller_al_2003, Muller_Grappin_2005, Mason_al_2008,
  Perez_Boldyrev_2009, Beresnyak_Lazarian_2009, Grappin_Muller_2010,
  Boldyrev_al_2011, Lee_al_2010, Chen_al_2011a, Beresnyak_2011,
  Perez_al_2012}.  These spectral indices are associated to the
different nature of the nonlinear interactions regulating the cascade and
the cascade rate.  Moreover, within the inertial range, a transition
between different regimes can occur \citep{Mininni_Pouquet_2007,
  Verdini_Grappin_2012}.  More sophisticated DNS, including other
physical processes like expansion effects \citep{Dong_al_2014},
Hall-MHD \citep[e.g.,][]{Matthaeus_al_2003, Gomez_al_2008,
  Shaikh_Shukla_2009, Shaikh_Zank_2009}, reduced Hall MHD
\citep{Gomez_al_2013}, gyrokinetic \citep{Howes_al_2008a,
  TenBarge_al_2013b}, and hybrid particle-in-cell (PIC) simulations
\citep{Vasquez_Markovskii_2012}, all produce spectral indices
consistent with $-5/3$. Anyway, the restricted width of the inertial
range prevents firm conclusions.  

As far as the small kinetic scales are concerned, DNS including proton
and electron physics return a qualitatively unified picture.  At
sub-proton scales, they reproduce an increase of the ratio of the
electric to magnetic power, together with a flattening of the electric
field spectrum \citep[e.g.,][]{Dmitruk_Matthaeus_2006a,
  Howes_al_2008a, Howes_al_2011,Servidio_al_2012,
  Gomez_al_2013,Perrone_al_2013,Parashar_al_2014,Passot_al_2014,
  Valentini_al_2014,Servidio_al_2015}, and a transition to a steeper
spectrum for the magnetic field power near the ion scales
\citep[e.g.,][]{Matthaeus_al_2003, Dmitruk_Matthaeus_2006a,
  Parashar_al_2010, Servidio_al_2012, Vasquez_Markovskii_2012,
  Rodriguez_al_2013, Valentini_al_2014}.  However, a unique spectral
index cannot be identified for the magnetic field spectrum at proton
scales.  Early works in Hall MHD \citep{Shaikh_Shukla_2009,
  Martin_al_2013}, Electron-MHD \citep{Biskamp_al_1999, Ng_al_2003,
  Cho_Lazarian_2004,Cho_Lazarian_2009, Shaikh_2009}, and gyrokinetic
\citep{Howes_al_2008a} reported a spectral index of $-7/3$ for the
magnetic field at sub-ion scales.  More recently, steeper spectra have
also been observed: a spectral index of $-2.8$ in gyrokinetic
\citep{Howes_al_2011, TenBarge_al_2013a,TenBarge_al_2013b} and finite
Larmor radius (FLR)-Landau fluid simulations \citep{Passot_al_2014},
or a $-8/3$ power law both in 3D electron-MHD
\citep{Meyrand_Galtier_2013} and in strong kinetic-Alfv\'en turbulence
\citep{Boldyrev_Perez_2012}. Magnetic spectral indices inbetween about
$-2.6$ and $-3$ have also been observed in full PIC simulations
\citep[e.g.,][]{Camporeale_Burgess_2011, Chang_al_2011, Wan_al_2012,
  Karimabadi_al_2013, Wu_al_2013}.

In situ measurements of the proton velocity
distribution functions show the presence of an ubiquitous temperature
anisotropy between the direction parallel and perpendicular to the
mean magnetic field \citep{Marsch_al_1982a, Hellinger_al_2006}, and a
non adiabatic evolution of the solar wind plasma during its expansion
\citep{Marsch_al_2004,Matteini_al_2007}, thus suggesting, as already
mentioned, an active role played by the turbulence in exchanging
energy between fields and particles.  Hybrid PIC simulations have
shown an overall (macroscopic) collisionless proton heating, with
signatures of a preferential proton heating in the perpendicular
direction with respect to the ambient mean magnetic field
\citep[e.~g.][]{Parashar_al_2009,
  Markovskii_al_2010,Markovskii_Vasquez_2011,
  Vasquez_Markovskii_2012,Verscharen_al_2012,Parashar_al_2014}.
Vlasov-Hybrid simulations suggest that non-Maxwellian kinetic effects,
such as temperature anisotropies, can be produced by the turbulence,
mostly concentrated in regions near and around the peaks of the
current density \citep{Servidio_al_2012, Valentini_al_2014,
  Perrone_al_2014b, Servidio_al_2015}.

In our previous work \citep{Franci_al_2015a} (named hereafter as
\citetalias{Franci_al_2015a}), we presented results from a
high-resolution hybrid (fluid electrons, PIC protons) two-dimensional
(2D) simulations of turbulence.  The spectra of various fluctuations
(magnetic, kinetic, density and electric field), along with the
magnetic compressibility and the non-dimensional ratio of the density
and the magnetic fluctuations, simultaneously matched several features
observed in the solar wind.  In particular, for the magnetic field we
showed that high-resolution hybrid simulations, although limited to a
2D geometry, are able to capture the nonlinear dynamics at fluid-like
MHD scales and at subproton scales, both within the same numerical
domain.  In this paper, we analyze in further detail the spectral
properties of several fields, also showing their stability with
time. Moreover, we investigate the shape of the electric field
spectrum, by estimating the separated contributions from different
terms in the generalized Ohm's law. Finally, we study the proton
temperature anisotropy and the proton heating, also quantifying the
dependence from the resistivity coefficient and the number of
particles-per-cell (ppc) employed in the simulations.

The paper is organized as follows: In Section~\ref{sec:setup}, we
describe the numerical setup employed, define the physical units and
normalizations in the code, and provide the parameters of our initial
conditions. In Section~\ref{sec:results}, we describe the results of
the performed simulations. In Section~\ref{sec:parameters}, we
validate such results, by investigating the importance of a
careful choice of some relevant numerical parameters.  Finally, in
Section~\ref{sec:conclusions}, we summarize the achievements of our
simulations and discuss them in the framework of both observational
and previous numerical and theoretical studies.

\section{Numerical setup and initial conditions}
\label{sec:setup}

We make use of a 2D hybrid code, where electrons are considered as a
massless, charge neutralizing fluid with a constant temperature,
whereas ions are described by a PIC model and are advanced by the
Boris' scheme, which requires the fields to be known at a half time
step ahead of the particle velocities. This is achieved by advancing
the current density to this time step with only one computational pass
through the particle data at each time step \citep{Matthews_1994}.

The characteristic spatial and temporal units used in this model are
the proton inertial length $d_p = c/\omega_p$, $\omega_p = ( 4 \pi n
e^2 / m_p )^{1/2}$ being the proton plasma frequency, and the
inverse proton gyrofrequency $\Omega^{-1}_p = (e B_0/m_p c
)^{-1}$, respectively. Magnetic fields are expressed in units of the
magnitude of the ambient magnetic field, i.e., $B_0$, while velocities 
are expressed in units of the Alfv\'en velocity, i.e., $v_A = B_0/ ( 4 \pi n m_p)^{1/2}$. 
The plasma beta for a given plasma species, protons ($p$) or electrons ($e$), is
$\beta_{p,e} = 8 \pi n K_B T_{p,e} / B^2_0$.  Quantities and symbols
used in these definitions are: the speed of light, $c$, the number
density, $n$, which is assumed to be equal for proton and electrons ($n_p
= n_e = n$), the magnitude of the electronic charge, $e$, the proton
mass, $m_p$, the Boltzmann's constant, $K_B$, 
and the proton and electron temperatures, $T_{p,e}$.

The 2D computational domain lies in the $(x,y)$ plane, while the
ambient magnetic field $\vect{B}_0$ is along the $z$-direction.
Accordingly, each field $\Psi$ will be decomposed in its perpendicular
(in-plane) component, $\Psi_\perp$, and its parallel (out-of-plane,
along $\vect{z}$) component, $\Psi_{\parallel}$, with respect to
$\vect{B}_0$. The only exceptions will be the proton beta and
temperature, for which $\perp$ and $\parallel$ will denote directions
with respect to the local magnetic field.

The adopted simulation box is a $2048^2$ square grid.
We tested different resolutions ($\Delta x = \Delta y =
0.5$, $0.25$ and $0.125 \, d_p$), and consequently different box sizes
($L_\text{box} = 1024, \, 512$ and $256 \, d_p$), as well as different
numbers of ppc, ranging from 500 to 8000 (see
Table~\ref{tab:modelslist}). The time step for the particle
advance is $\Delta t = 0.025 \, \Omega_p^{-1}$, whereas the magnetic field
$\vect{B}$ is advanced with a smaller time step, $\Delta t_\text{B} =
\Delta t/10$.  
\begin{table}[!t]
\begin{tabularx}{0.48\textwidth}{Y|YY|Y|Y} 
\hline \hline
    & $\Delta x$ & $L_{\text{box}}$ & $\eta$            &     \\
Run & $(d_p)$    & $(d_p)$          & $(4\pi/\omega_p)$ & ppc \\
\hline \hline
A & 0.125 & 256 & $5 \times 10^{-4}$ & 8000 \\
\hline
B & 0.125 & 256 & $5 \times 10^{-4}$ & 4000 \\
C & 0.125 & 256 & $5 \times 10^{-4}$ & 2000 \\
D & 0.125 & 256 & $5 \times 10^{-4}$ & 1000 \\
E & 0.125 & 256 & $5 \times 10^{-4}$ &  500 \\
\hline
F & 0.125 & 256 & $1 \times 10^{-4}$ & 8000 \\
G & 0.125 & 256 & $1 \times 10^{-3}$ & 8000 \\
\hline
H & 0.250 & 512  & $1 \times 10^{-3}$ & 8000 \\ 
I & 0.500 & 1024 & $2 \times 10^{-3}$ & 8000 \\ 
\hline \hline
\end{tabularx}
\caption{List of simulations and their relevant parameters}
\label{tab:modelslist}
\end{table}

Initially, we assume a uniform number density $n = 1$, a proton parallel beta
$\beta_{p\parallel} = 0.5$, and a temperature anisotropy $A_p =
T_{p\perp} / T_{p\parallel} = 1$.  Electrons are isotropic, with
$\beta_e = 0.5$.  We add an initial spectrum of magnetic and velocity
fluctuations in the $(x,y)$ plane, composed of modes with $-0.2 <
k_{x,y} < 0.2$ in each direction and random phases. These initial
fluctuations are characterized by energy equipartition and vanishing
correlation between kinetic and magnetic fluctuations, and their
global amplitude is $B^{rms} \sim 0.24$. Hereafter, 
\begin{equation}
\label{eq:rms}
\Psi^{rms} = (\langle \Psi^2 \rangle - \langle \Psi \rangle^2)^{1/2}
\end{equation}
will denote the root mean square value (rms) of a quantity $\Psi$,
with $\langle\Psi\rangle$ being its space-averaged value 
over the whole 2D simulation domain.  
The initial magnetic fluctuations can be expressed in the form
\begin{equation}
\label{eq:bfluctuations}
\begin{split}
\vect{B}_{\perp}(x,\,y) &= \frac{1}{2} \sum_{k_x,k_y} [
  \vect{B}_{\perp}(k_x,\,k_y) \\ & \times \exp(i(k_x x + k_y y
  + \phi(k_x,\,k_y))) + \text{c.c.}].
\end{split}
\end{equation}
The initial bulk velocity fluctuations $\vect{u}_\perp(t=0)$ are
assumed to have the same form as in Eq.~\eqref{eq:bfluctuations}, with
different random phases.

We introduce two dimensionless quantities, i.e., the normalized 
cross helicity, $\sigma_\text{C}$, and the normalized residual
energy, $\sigma_R$:  
\begin{align}
  \sigma_\text{C}(x,y) &= \frac{2 \; \vect{u} \cdot
    \vect{B}}{|\vect{u}|^2 + |\vect{B}|^2},
\label{eq:sigmaC} \\
  \sigma_\text{R}(x,y) &= \frac{|\vect{u}|^2 -
    |\vect{B}|^2}{|\vect{u}|^2 + |\vect{B}|^2},
\label{eq:sigmaR}
\end{align}
which define the two-dimensional geometry of the fluctuations.  
With the initial conditions we chose for the initial
magnetic and bulk velocity fluctuations, $\sigma_\text{C}$ and
$\sigma_\text{R}$ are both statistically very close to zero, even
though they are not actually zero anywhere in the $(x,\,y)$ plane.

A non-zero resistivity has been introduced in order to
guarantee a satisfactory conservation of the total energy, with no
claim to model any realistic physical process. Its value has been
empirically fine-tuned by running different simulations 
(see Table~\ref{tab:modelslist}). Further details about this point
will be provided in Sec.~\ref{sec:parameters}.

\section{Results}
\label{sec:results}
We performed nine different simulations.  Their main parameters are
listed in Table~\ref{tab:modelslist}. A label is assigned to each run
in the first column, while in the other columns we report, from left to
right: the spatial resolution, $\Delta x$ ($= \Delta y$), the length
of the simulation box, $L_{\text{box}}$, the value of the resistivity
coefficient, $\eta$, and the number of ppc.

Run A employs the best spatial resolution, ${\Delta x = \Delta y =
  0.125 \, d_p}$, and the highest number of particles, i.e., 8000 ppc,
corresponding to more than $3 \times 10^{10}$ particles in the whole
simulation domain.  The resistive coefficient has been fine-tuned and
set to the value $\eta=5 \times 10^{-4}$, in units of $4\pi
\omega_p^{-1}$.  In Subection~\ref{subsec:runA} and
\ref{subsec:spectra} we will provide a detailed and quantitative
analysis of the data produced by this run. The remaining simulations were
performed in order to validate these results and to investigate the
effects of the number of ppc (Runs B-E, see
Table~\ref{tab:modelslist}), the resistivity (Runs F-G), and the
spatial resolution (Runs H-I). Their results will be discussed later,
in Section~\ref{sec:parameters}.

\subsection{Temporal and spatial evolution}
\label{subsec:runA}

\begin{figure}
\centering
\includegraphics[width=0.45\textwidth]{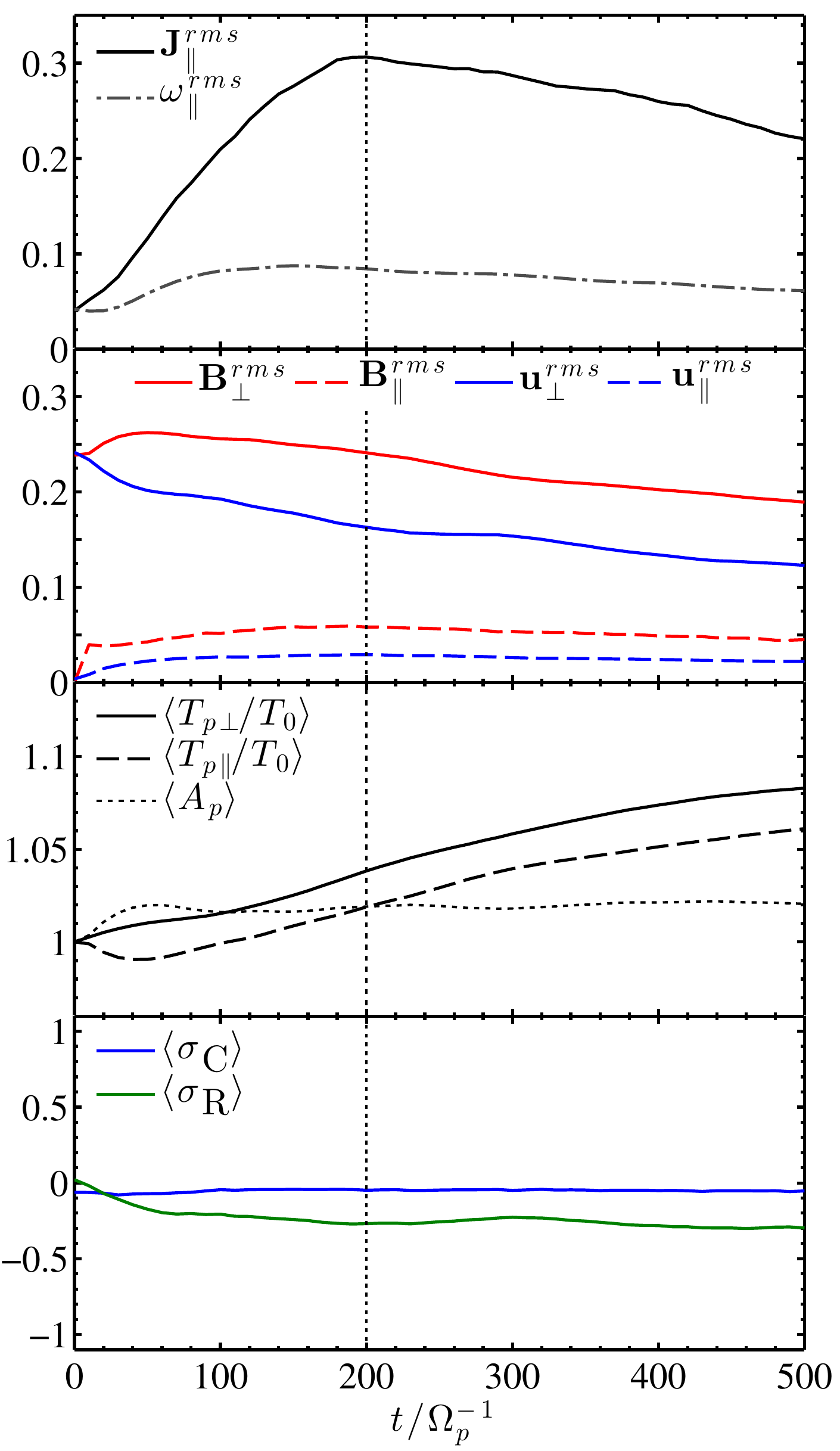}\\
\caption{Time evolution of space-averaged quantities.  From top to
  bottom: the rms out-of-plane current density,
  $\vect{J}_\parallel$, and the rms out-of-plane vorticity,
  $\vect{\omega}_{\parallel}$ (\textit{first panel}); the rms 
  perpendicular $\vect{B}_\perp$ and parallel $\vect{B}_\parallel$
  magnetic fluctuations, and the rms perpendicular $\vect{u}_\perp$ and
  parallel $\vect{u}_\parallel$ velocity fluctuations (\textit{second
    panel}); the mean values of the normalized perpendicular and
  parallel proton temperatures, $T_{p\perp}/T_0$ and
  $T_{p\parallel}/T_0$, and of the proton temperature anisotropy,
  $A_{p}$ (\textit{third panel}); the mean values of the normalized
  cross helicity, $\sigma_\text{C}$, and of the normalized residual energy,
  $\sigma_\text{R}$ (\textit{fourth panel}). In all panels, a vertical
  black dotted line marks the time of the maximum turbulent activity,
  i.e., $t = 200 \, \Omega_p^{-1}$.}
\label{fig:time_evolutions}
\end{figure}
\begin{figure*}
\centering 
\includegraphics[width=\textwidth]{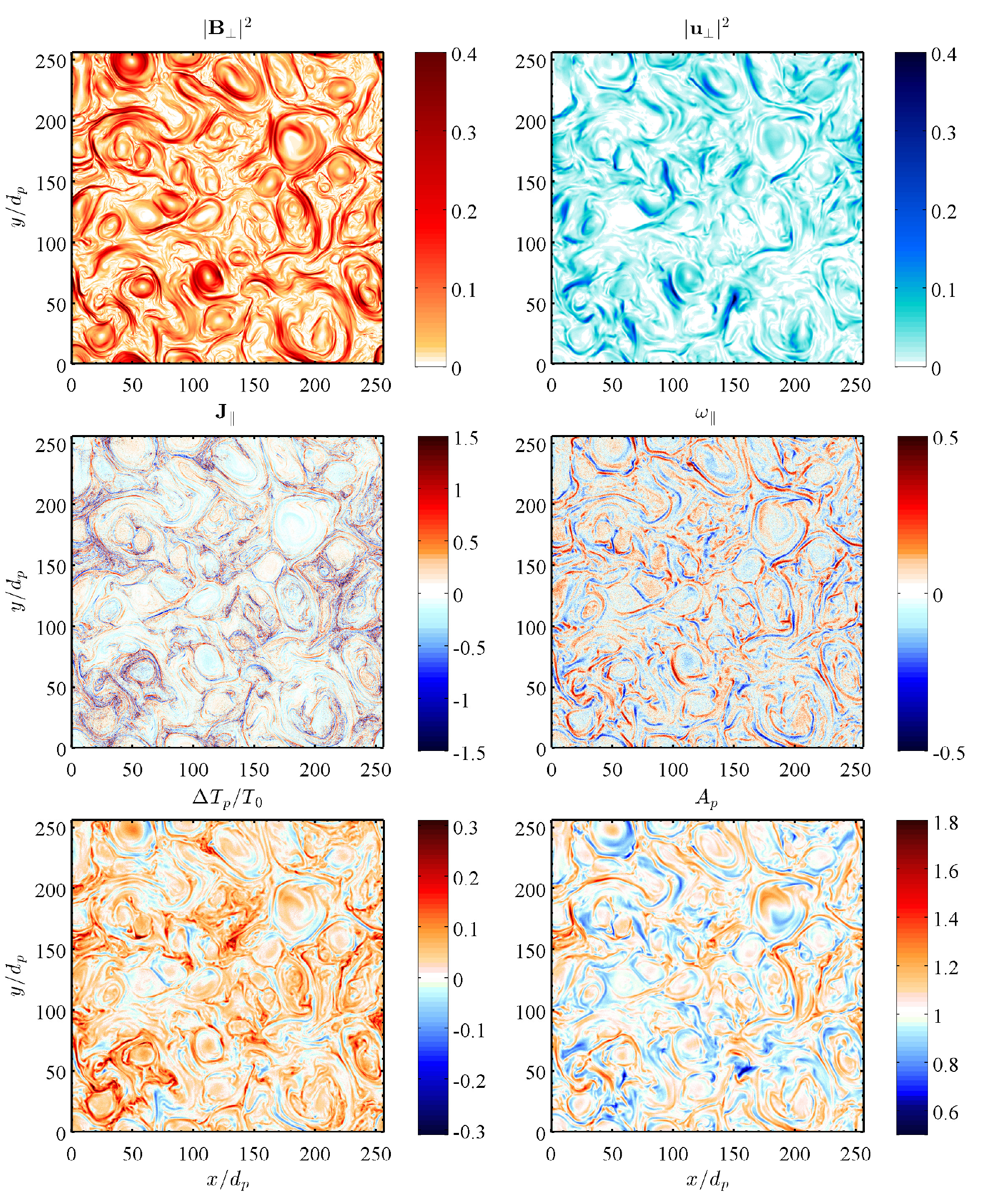}
\caption{Contour plots of six different quantities on the $(x,y)$
  plane at $t = 200 \, \Omega_p^{-1}$: the amplitude of the perpendicular
  magnetic fluctuations, $\vect{B}_\perp^2$ (\textit{upper-left panel}),
  and of the perpendicular velocity fluctuations, $\vect{u}_\perp^2$
  (\textit{upper-right panel}), the out-of-plane current density,
  $\vect{J}_\parallel$ (\textit{middle-left panel}), and vorticity,
  $\vect{\omega}_\parallel$ (\textit{middle-right panel}), the proton
  temperature variation normalized to the initial temperature, $\Delta
  T_p/T_0$ (\textit{bottom-left panel}), and the proton temperature
  anisotropy, $A_{p}$ (\textit{bottom-right panel}).}
\label{fig:turbulence}
\end{figure*}

In Fig.~\ref{fig:time_evolutions}, the time evolution of a few
quantities is shown up to $500 \,\Omega_p^{-1}$.  The initial
non-linear time associated to the maximum injection scale, i.e.,
$k^{\textrm{inj}} \sim 0.2 \, d_p^{-1}$, can be estimated as
$t_{\textrm{NL}} \sim [ k^{\textrm{inj}} \delta u_{ k^{\textrm{inj}}
}]^{-1} \sim 20 \,\Omega_p^{-1}$, and corresponds to the minor ticks
of the $x$ axis.  The total length of the simulation is 
approximately 25 $t_{\textrm{NL}}$. 

In the first panel, from top to bottom, we report the rms out-of-plane
current density, $\vect{J}_{\parallel}$, and the rms
out-of-plane vorticity, $\vect{\omega}_{\parallel}$. The current
increases quite rapidly, attains its maximum value just before $t \sim
200 \, \Omega_p^{-1}$ and then slowly decreases. Since it represents a
good indicator of the level of turbulent
activity~\citep{Mininni_Pouquet_2009}, we choose to perform a detailed
analysis at $t \sim 200 \, \Omega_p^{-1}$, when the turbulence is
expected to be fully developed.  A vertical black dotted line marks
this time in all four panels. The vorticity also increases quite
rapidly, reaching an earlier and lower maximum value, and then it
decreases extremely slowly.

The second panel of Fig.~\ref{fig:time_evolutions} shows the rms  
perpendicular $\vect{B}_{\perp}$ and parallel
$\vect{B}_{\parallel}$ magnetic fluctuations (red lines), and the rms
perpendicular $\vect{u}_{\perp}$ and parallel $\vect{u}_{\parallel}$
velocity fluctuations (blue lines).  Perpendicular and parallel
components are drawn with solid and dashed lines respectively.
$\vect{B}_{\perp}$ exhibits a small increase until $t \sim 40 \,
\Omega_p^{-1}$,
and then it decreases quite smoothly. On the other hand,
$\vect{u}_{\perp}$ decreases with a similar trend, but without showing
any initial growth.  This indicates that the turbulence is fed by the
perpendicular components of both the magnetic and the velocity fluctuations,
whose energy decreases slowly and sustains the cascade for the whole
evolution.  Contextually, the parallel components of both the magnetic and
the velocity fluctuations rapidly originate from compressive effects,
remaining much smaller than their perpendicular counterparts
throughout the simulation.

In the third panel of Fig.~\ref{fig:time_evolutions}, we report the
space-averaged parallel and perpendicular proton temperatures,
normalized to the initial value $T_0$, $\langle
T_{p\parallel}/T_0\rangle$ and $\langle T_{p\perp}/T_0\rangle$
respectively, together with the space-averaged proton temperature
anisotropy, $\langle A_{p}\rangle = \langle
T_{p\perp}/T_{p\parallel}\rangle$. We recall here that $T_{\parallel}$ and
$T_{\perp}$ are defined with respect to the local magnetic field.
Since $\langle A_{p}\rangle = 1$ is imposed at $t=0$, $T_{p\parallel}$
and $T_{p\perp}$ share the same initial value, $T_0$. In the very first
part of the evolution, the former shows a little and sudden decrease,
after which both increase with almost the same rate. The parallel and
perpendicular energy gains, at the end of the simulation, i.e., at $t = 500
\, \Omega_p^{-1}$, are approximately 6\% and 8\% respectively. This
small excess of perpendicular energy quickly arises within $\sim 2 \,
t_{\textrm{NL}}$, and it is preserved throughout the simulation,
with the temperature anisotropy reaching a value $\langle A_{p}
\rangle \sim 1.02$ in correspondence of the maximum turbulent activity,
and then remaining quite constant until the end of the simulation.  A
detailed discussion about the proton heating will be further provided in
Subsection~\ref{subsec:heating}.

Lastly, in the bottom panel of Fig.~\ref{fig:time_evolutions}, we show
the space-averaged values of the normalized cross helicity,
$\sigma_C$, and of the residual energy, $\sigma_R$, (see
Eq.~\eqref{eq:sigmaC} and~\eqref{eq:sigmaR}). The former is very close
to zero at the beginning of the simulation (as a result of the
initially imposed random phases spectra), and it tends to maintain this
value until the end.  The latter, instead, decreases from zero to about
$-0.3$ in very few non-linear times $t_{\textrm{NL}}$, showing a
global excess of the magnetic energy over the kinetic energy.  These
asymptotic values are reached very quickly, as a consequence of the
relaxation from the initial random relative orientation of the velocity
and the magnetic fluctuations towards a strongly aligned state.  Despite
the steady time evolution of their space-averaged values, both
$\sigma_C$ and $\sigma_R$ appear very patchy when looking at the
spatial distribution throughout the 2D computational domain (not
shown), exhibiting quite a wide excursion from $-1$ to $1$ between
different albeit close regions.

\begin{figure}[t!]
\centering
\includegraphics[width=0.48\textwidth]{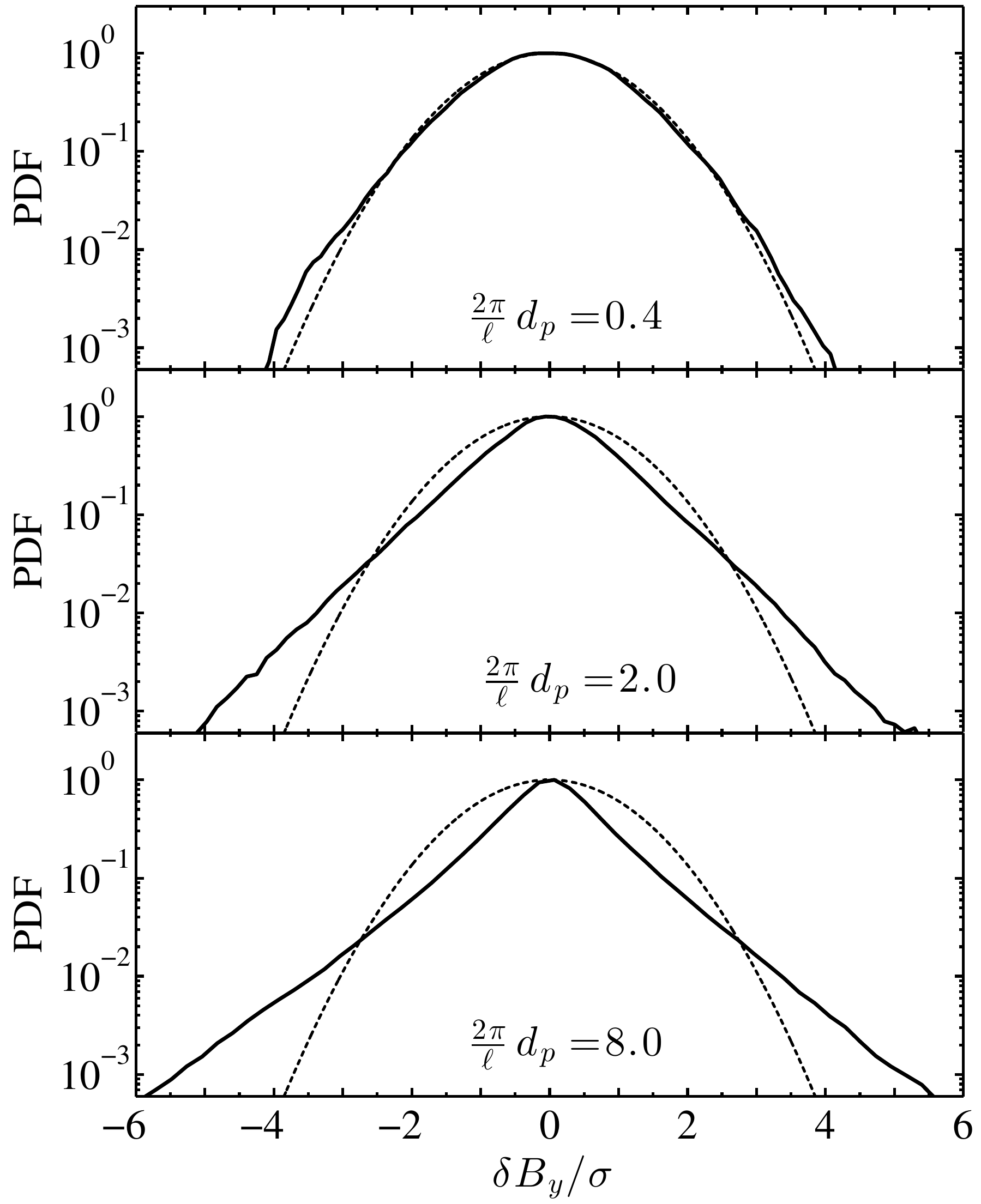}\\ 
\includegraphics[width=0.48\textwidth]{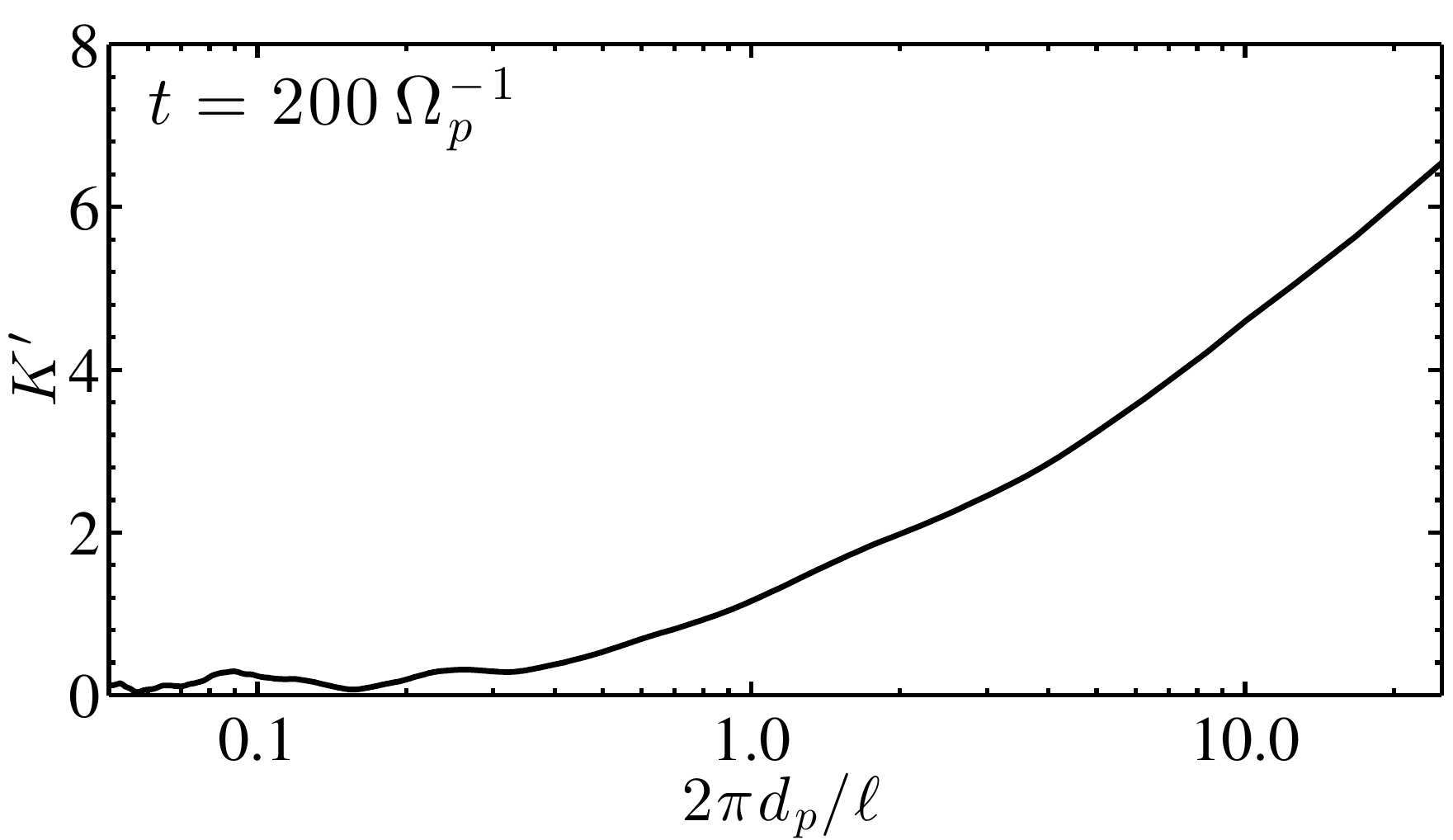}\\
\caption{Top panel: PDFs of the increments of the perpendicular
  magnetic field component $B_y$ along $x$ at $t = 200 \,
  \Omega_p^{-1}$, corresponding to $2\pi d_p/ \ell = 0.4$, $2\pi d_p/
  \ell = 2$ and $2\pi d_p/ \ell = 8$ (from top to bottom). In each
  panel, a Gaussian function with the same variance is plotted with a
  dashed line as a reference. Bottom panel: Excess kurtosis of the
  same quantity, computed at the same time.}
\label{fig:PDFs_kurtosis}
\end{figure}

Summarizing the time evolution of all the above-mentioned quantities,
we can divide the evolution of the system in three different stages:
\begin{enumerate}
\item a rapid re-adjustment and relaxation of the initial conditions,
  occurring within $t \lesssim 40 \, \Omega_p^{-1} \sim 2 \,
  t_\textrm{NL}$
\item the onset of a turbulent cascade, fed by the perpendicular
  magnetic and velocity fluctuations, involving larger and larger
  scales on times of the order of $t \sim 200 \, \Omega_p^{-1} \sim 10
  \, t_\textrm{NL}$
\item a deacaying phase with slow and smooth variations of all rms
  quantities, during which the turbulence is fully developed and
  further sustained until at least $t \sim 500 \, \Omega_p^{-1}$,
  corresponding to $\sim 25 \, t_\textrm{NL}$.
\end{enumerate}
 
Fig.~\ref{fig:turbulence} shows isocontours of six different
quantities in the whole simulation domain, all computed at $t = 200 \,
\Omega_p^{-1}$.  In the upper-left panel, we report the magnitude of
the perpendicular magnetic fluctuations, $|\vect{B}_\perp|^2$, showing
the presence of coherent structures in the magnetic field, i.e.,
vortices and magnetic islands, embedded in a much more chaotic
environment where stretched and twisted shapes emerge. In the
upper-right panel, the magnitude of the perpendicular velocity
fluctuations, $|\vect{u}_\perp|^2$, is shown to exhibit qualitatively
the same kind of structures, but with lower intensity and much lower 
gradients. In some regions, high values of $|\vect{u}_\perp|^2$
correspond to high values of $|\vect{B}_\perp|^2$, while in other
regions the opposite situation holds. In the middle-left panel, we show
the out-of-plane current density, $J_\parallel = (\nabla \times
\vect{B})_\parallel$.  Many 
thin current sheets form, since the very first phase of the evolution,
mostly around and in-between vortices. Once formed, each current sheet
is quickly disrupted into smaller and smaller pieces, contributing to the
generation of smaller-scale structures. At the time of maximum turbulent
activity, this results in the articulated pattern shown here.
In the middle-right panel, the out-of-plane vorticity,
$\vect{\omega}_\parallel = (\nabla \times \vect{u})_\parallel$, is
shown.  It exhibits a structure similar to $\vect{J}_\parallel$, with
many thin layers, whose shape is however much more defined and clean in
respect to $J_\parallel$. Peaks of $\vect{\omega}_\parallel$ and
peaks of $\vect{J}_\parallel$ occupy approximately the same regions.
In the bottom left panel, we report the proton temperature variation in respect
to the initial proton temperature, $\Delta T_p/T_0 = (T_p - T_0)/T_0$, 
where $T_p = (2\, T_{p\perp} + T_{p\parallel})/3$ is the average proton temperature
measured at $t = 200 \, \Omega_p^{-1}$.
Regions where $\Delta T_p$ is locally both negative or positive are
clearly present, and a resulting global proton temperature enhancement
can be observed, as already inferred from Fig.~\ref{fig:time_evolutions}.  
Interestingly, areas where a proton temperature enhancement occurs are
located in the vicinity of current sheets (cf.,~\citep{Servidio_al_2012}). 
A more detailed analysis shows that strong currents exhibit a complex
evolution, which involves splitting/dissociation and leads to a relevant
proton energization.

In the bottom right panel, the proton temperature
anisotropy, $A_{p}$, is shown.  We observe a wide excursion between very
close areas, the perpendicular proton temperature $T_{p\perp}$ ranging
from about half and almost twice the parallel one. Therefore, there is a
strong local reshaping of particle distributions, leading to both
perpendicular and parallel anisotropies \citep{Servidio_al_2014a}.
Nevertheless, as inferable from Fig.~\ref{fig:time_evolutions}, the
relative difference between $\langle T_{p\perp}\rangle$ and $\langle
T_{p\parallel}\rangle$ is about 2\% at $t = 200 \, \Omega_p^{-1}$,
meaning that globally no preferential enhancement along the perpendicular
or parallel direction is achieved.  

The small-scale coherent structures which have emerged by the time of
maximum turbulent activity, already observed in
Fig.~\ref{fig:turbulence}, can be related to the phenomenon of
intermittency, since they are able to induce departures from
self-similarity and enhanced dissipation. In order to look for
intermittency in our data, we examine the
non-Gaussian behavior of the probability density function (PDF) of a
MHD primitive variable. In particular, we compute
the PDFs at $t = 200 \, \Omega_p^{-1}$ by taking increments of one of
the perpendicular magnetic field components, i.e., $\vect{B}_y$, along
the other perpendicular direction, i.e., $x$, for three different
spatial separations, $\ell$. In the three top panels of
Fig.~\ref{fig:PDFs_kurtosis}, we show three PDFs, computed for $2\pi
d_p/\ell = 0.4$, which is approximatively in the middle of the
inertial range (upper panel), $2\pi d_p/\ell = 2$, which is the scale
corresponding to the ion spectral break (middle panel) and $2\pi
d_p/\ell = 8$, which is well inside the kinetic range (bottom panel).
A Gaussian function with the same variance is plotted with a dashed
line in each panel as a reference.  The distribution of magnetic
fluctuations is clearly different at different scales: it is closer to
a normal distribution at very large scales, it shows a significant
deviation at intermediate scales, and it displays very extended tails
at small scales. In order to
quantify the level of intermittency, we compute the fourth central
moment $K$ (or kurtosis) of the distributions.  In the bottom panel of
Fig.~\ref{fig:PDFs_kurtosis}, we show the excess kurtosis $K' = K - 3$,
computed from the increments of $\vect{B}_y$ along $x$, as a function 
of $2\pi d_p/\ell$, again at $t = 200 \, \Omega_p^{-1}$.  This quantity is
clearly very close to zero up to the injection scale, i.e., $k_{\perp}\,d_p 
\lesssim 0.2$, and the it  steadily increases through the inertial 
range and down to sub-proton scales.
\begin{figure}[t!]
\centering
\includegraphics[width=0.47\textwidth]{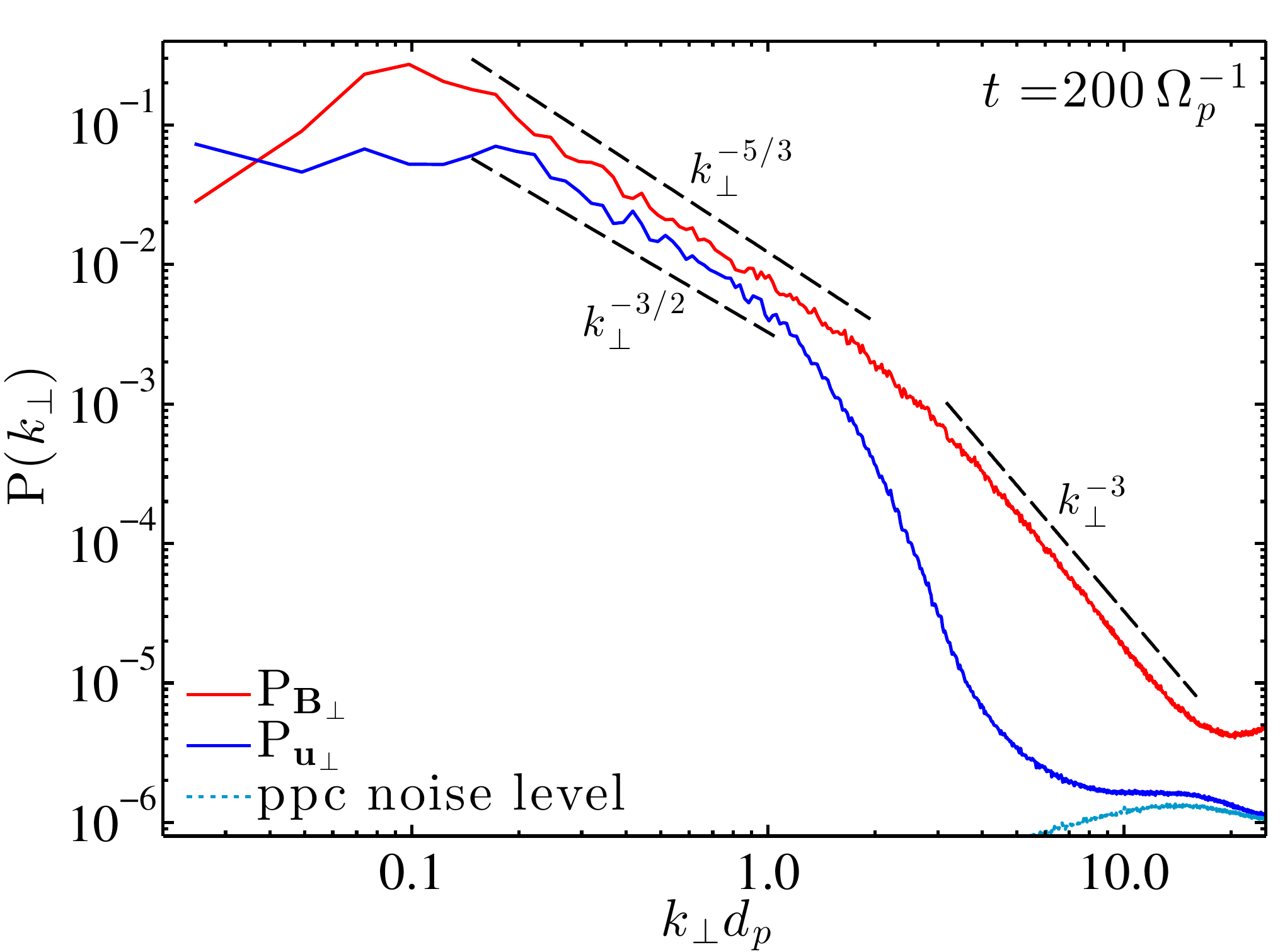}
\caption{Power spectra of the perpendicular magnetic and velocity
  fluctuations, $\vect{B}_{\perp}$ (red solid line) and
  $\vect{u}_{\perp}$ (blue solid line), respectively. Power laws with
  different spectral indices are additionally shown in black dashed
  lines as a reference.}
\label{fig:spectra}
\end{figure}
\begin{figure}[t!]
\centering
\includegraphics[width=0.47\textwidth]{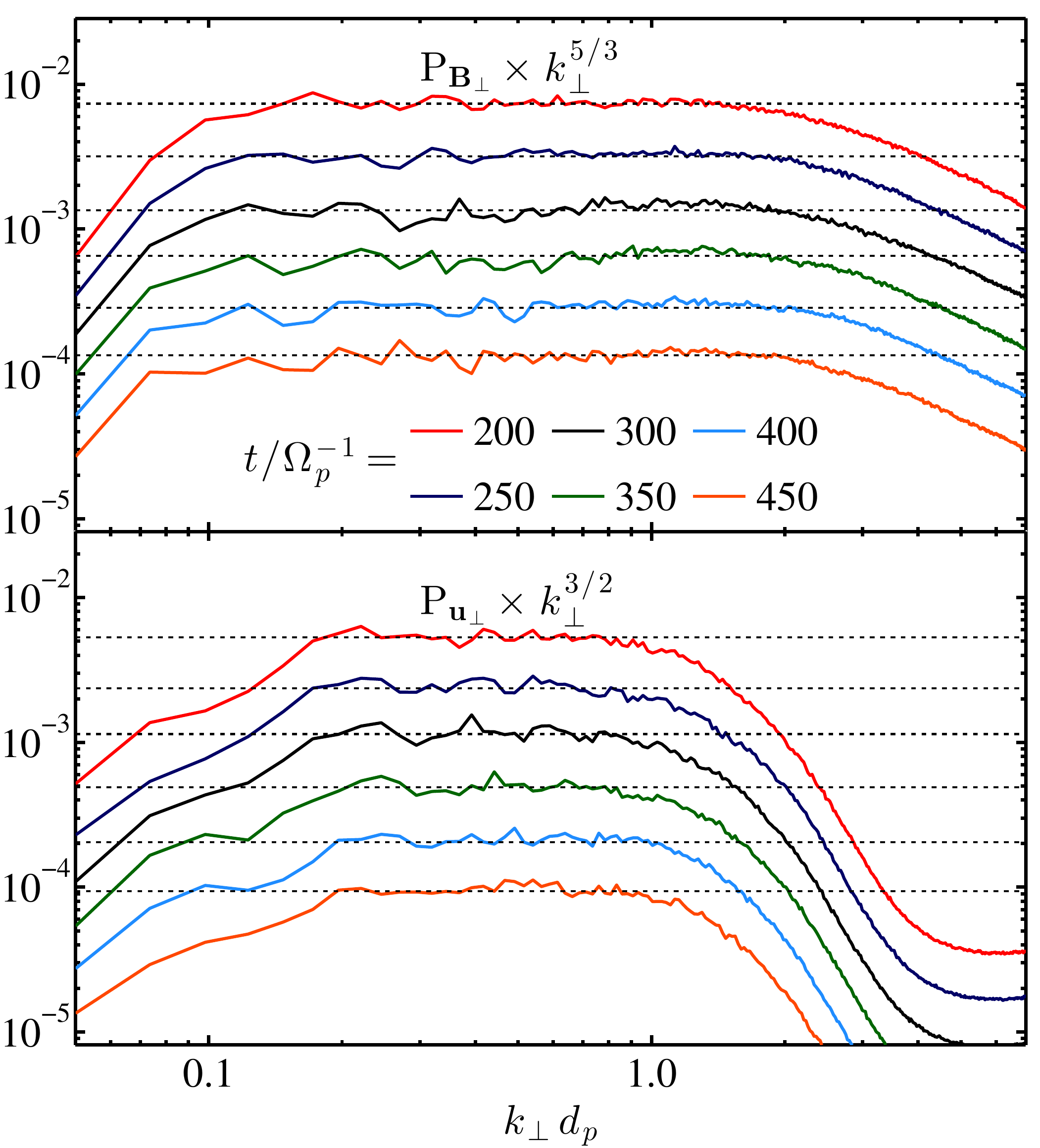}
\caption{Top panel: Power spectra of the perpendicular magnetic
  fluctuations, $\vect{B}_{\perp}$, compensated by
  $k_\perp^{5/3}$. Each of them is the time average in the interval
  $[\tilde{t} - 20 \, \Omega_p^{-1} ,\, \tilde{t} + 20 \,
    \Omega_p^{-1}]$, where $\tilde{t}$ is the time reported in the
  legend. Note that all of them have been suitably rescaled for the
  sake of clarity, so that they did not overlapped each other.
  Horizontal dotted black lines are additionally shown.  Bottom panel:
  The same as in the upper panel but for the perpendicular velocity
  fluctuations, $\vect{u}_{\perp}$, compensated by $k_\perp^{3/2}$.}
\label{fig:spectra_comp}
\end{figure}

\subsection{Spectral properties}
\label{subsec:spectra}
Since the small-scale structures shown in Fig.~\ref{fig:turbulence}
exhibit random orientations, and therefore the two-dimensional spectra
of all fluctuations can be assumed statistically isotropic, we can
perform a quantitative analysis of the turbulent cascade by computing
the omnidirectional spectra. These are defined as
\begin{eqnarray}
P_\Psi(k_\bot)\equiv\delta
\Psi^2(k_\bot)/k_\bot=\sum_{|\vect{k_\bot}|=k_\bot}\hat{\Psi}^2_{2D}(\vect{k}_\bot),
\end{eqnarray}
where $\hat{\Psi}$ are the Fourier coefficients of a given quantity $\Psi$,
and $\delta \Psi(k_\bot)$ is the amplitude of the fluctuation $\Psi$ at the
scale $k_\bot$.

In Fig.~\ref{fig:spectra}, we show the spectra of the perpendicular
magnetic and velocity fluctuations, drawn with red and blue solid
lines respectively, at $t = 200 \, \Omega_p^{-1}$.  We clearly observe
two power-law ranges, separated by a smooth spectral break at a scale
of the order of the the proton inertial length, $k_\perp\,d_p \sim
2$. In \citetalias{Franci_al_2015a}, we showed the spectra of the
total magnetic and velocity fluctuations, which exhibit a very similar
behavior, since the perpendicular components are the dominant ones for
both field.

In particular, in the inertial range the spectrum of the perpendicular
magnetic fluctuations follows a Kolmogorov $k_{\perp}^{-5/3}$
power-law scaling over a full decade in wavenumber,
approximately between $k_\perp d_p = 0.1$ and $2$.  Simultaneously,
the perpendicular proton bulk velocity fluctuations exhibit a less
steep slope, with an Iroshnikov-Kraichnan scaling of
$k_{\perp}^{-3/2}$, over a little less than a decade, in the range $0.2 \lesssim
k_\perp d_p \lesssim 1$. Moreover, an excess of magnetic energy over
kinetic energy is observed, coherently with the negative value of the
normalized residual energy $\sigma_{\textrm{R}}$ already shown in the
bottom panel of Fig.~\ref{fig:time_evolutions}.

At kinetic scales, the spectra of both fields steepen, due to the
presence of both kinetic and dissipative (resistive) effects.  The
spectrum of $\vect{u}_\perp$ quickly drops with an exponential trend
above $k_\perp d_p \sim 1$, until it clearly saturates to the noise
level due to the finte number of ppc, corresponding to the spectrum at
$t=0$.  The spectrum of the magnetic fluctuations, on the contrary,
continue to follow a power-law scaling also at sub-proton scales,
although with a steeper spectral index, of the order of $-3$.  For
$k_\perp d_p \gtrsim 10$, $P_{\vect{B}_\perp}$ does not show an
exponential damping, as one would expect for resistive dissipation,
but a small increase instead, since the adopted resistive coefficient
is slightly smaller than the optimal value.

As discussed, the maximum level of turbulent activity occurs at $t
\sim 200 \, \Omega_p^{-1}$, which is about ten times the initial
nominal nonlinear time $t_{\textrm{NL}}$.  This can be explained with
the fact that at $t = 0$ we inject energy through several modes within
the range $[k_0,k^{\textrm{inj}}]$, where $k_0$ is the largest scale
corresponding to the computational box size, i.e., $k_0 = 2 \pi /
(256\,d_p) \sim 0.025 \,d_p^{-1}$.  The nominal nonlinear time
$t_{\textrm{NL}}|_{t=0}$ is different for each mode, being longer for
lower $k$-vectors. As the system evolves, the injection scale gets
larger and larger and most of the initial modes are involved in the
development of the turbulent cascade at $t = 200 \, \Omega_p^{-1}$.
Since modes with lower $k$s keep feeding energy at large scales even
afterwards, we expect turbulence to be still sustained also at later
times.

In Fig.~\ref{fig:spectra_comp} we show the spectra of the
perpendicular magnetic fluctuations, compensated by $k_\perp^{5/3}$
(top panel), and the spectra of the perpendicular velocity fluctuations,
compensated by $k_\perp^{3/2}$ (bottom panel), computed at regular intervals of
$50\, \Omega_p^{-1}$, from the maximum of turbulent activity to almost
the end of the simulation. Here, the spectrum at a given time
$\tilde{t}$ is the time-average between five different
spectra corresponding to $\tilde{t}$, $\tilde{t} \pm 10 \,
\Omega_p^{-1}$ and $\tilde{t} \pm 20 \, \Omega_p^{-1}$. The power-law
scaling for both the magnetic and the velocity fluctuations are very well
maintained, over about the same range, at all times $t > 200 \,
\Omega_p^{-1}$, indicating that the turbulence decays in a self-similar
way.  Note that spectra corresponding to different times have
been slightly shifted along the vertical axis, in order to avoid overlapping. 

\begin{figure}[t!]
\centering
\includegraphics[width=0.48\textwidth]{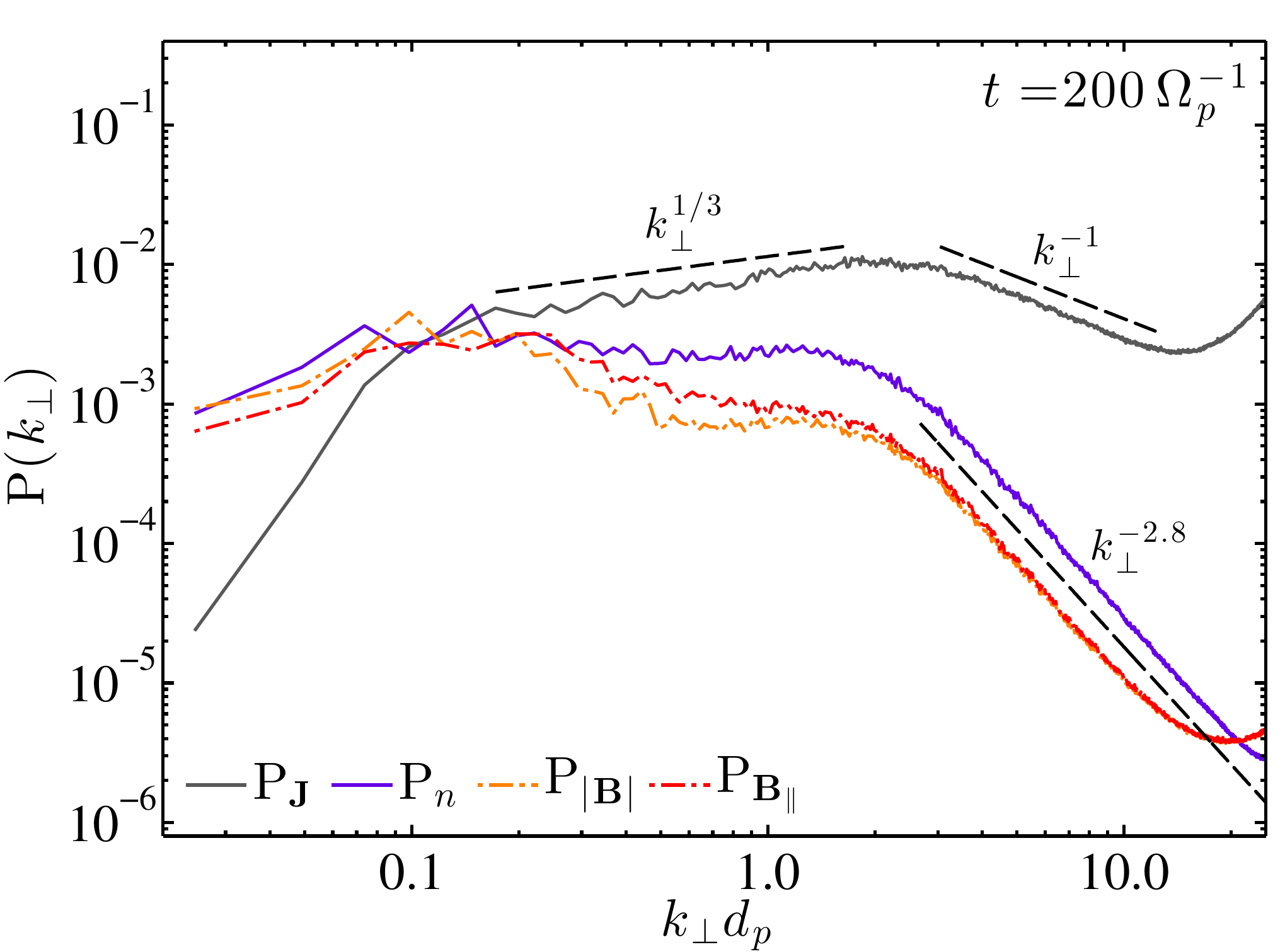}
\caption{Spectra of the total current density, $\vect{J}$ (grey line),
  of the density fluctuations, $n$ (purple line), of the magnitude of the
  magnetic field, $|\vect{B}|$ (orange line), and of its parallel
  component, $\vect{B}_\parallel$ (red dot-dashed line).  Power laws
  with different spectral indices are additionally drawn in black dashed
  lines, as a reference.}
\label{fig:spectra2}
\end{figure}
In Fig.~\ref{fig:spectra2}, we show the spectra of the magnitude of
the magnetic field, $|\vect{B}|$ (orange), of its parallel component,
$\vect{B}_\parallel$ (red dot-dashed), of the density fluctuations,
$n$ (purple) and of the total current density, $\vect{J}$ (grey).  
The density and the parallel magnetic fluctuations are
strongly coupled beyond $k_\perp d_p \sim 2$. In the MHD range, they
exhibit a flat spectrum, which is approximately an order of magnitude
smaller than the one of the perpendicular magnetic
fluctuations. Therefore, the large-scale activity has little
contribution from compressible fluctuations -- although they can still
play a significant role in the dynamics of the out-of-plane components
-- and the magnetic compressibility, i.e., the ratio of parallel to
total magnetic fluctuations, is also negligible at small $k$s. Both
spectra steepen at sub-proton scales, following a clean power-law
scaling with a spectral index of $-2.8$. Note that their relative power
level with respect to other fields' spectra increases, with
$P_{\vect{B}_\parallel}$ (and also $P_{|\vect{B}|}$) becoming
comparable with the spectrum of the perpendicular component,
$\vect{B}_\perp$ (cf. Fig.~\ref{fig:spectra}).

The spectral shape of the current density, $\vect{J}$, can be understood
by recalling that $\vect{J} = \bnabla \times \vect{B}$.  A simple
order-of-magnitude estimate of its perpendicular and parallel
components gives $\vect{J}_\perp \propto k_\perp \vect{B}_\parallel$
and $\vect{J}_\parallel \propto k_\perp \vect{B}_\perp$,
respectively. Therefore, in the inertial range, where the magnetic
activity is dominated by the perpendicular fluctuations, the spectrum
of $\vect{J}$ is determined by its parallel component
$\vect{J}_\parallel$ and this results in the observed spectral index of
$+1/3$, since $P_{\vect{J}_\parallel} \propto k^2_\perp
P_{\vect{B}_\perp}$, with $P_{\vect{B}_\perp} \propto k^{-5/3}_\perp$.
On the contrary, as already discussed, $P_{\vect{B}_\parallel}$ and
$P_{\vect{B}_\perp}$ are comparable at sub-proton scales, both showing
a power-law scaling, with spectral indices of $-2.8$ and $-3$,
respectively.  The corresponding components, $\vect{J}_\parallel$ and
$\vect{J}_\perp$, are of the same order and exhibit a similar scaling,
therefore $P_{\vect{J}} \sim P_{\vect{J}_\parallel} \sim
P_{\vect{J}_\perp}$, and the corresponding scaling is in-between 
$\propto k^{-0.8}$ and $\propto k^{-1}$.
The change in the spectral slope of $P_{\vect{J}}$ at $k_\perp d_p\sim 2$ 
provides a further evidence of a spectral break in the magnetic field
spectrum at proton scales.
\begin{figure}[t!]
\centering
\includegraphics[width=0.48\textwidth]{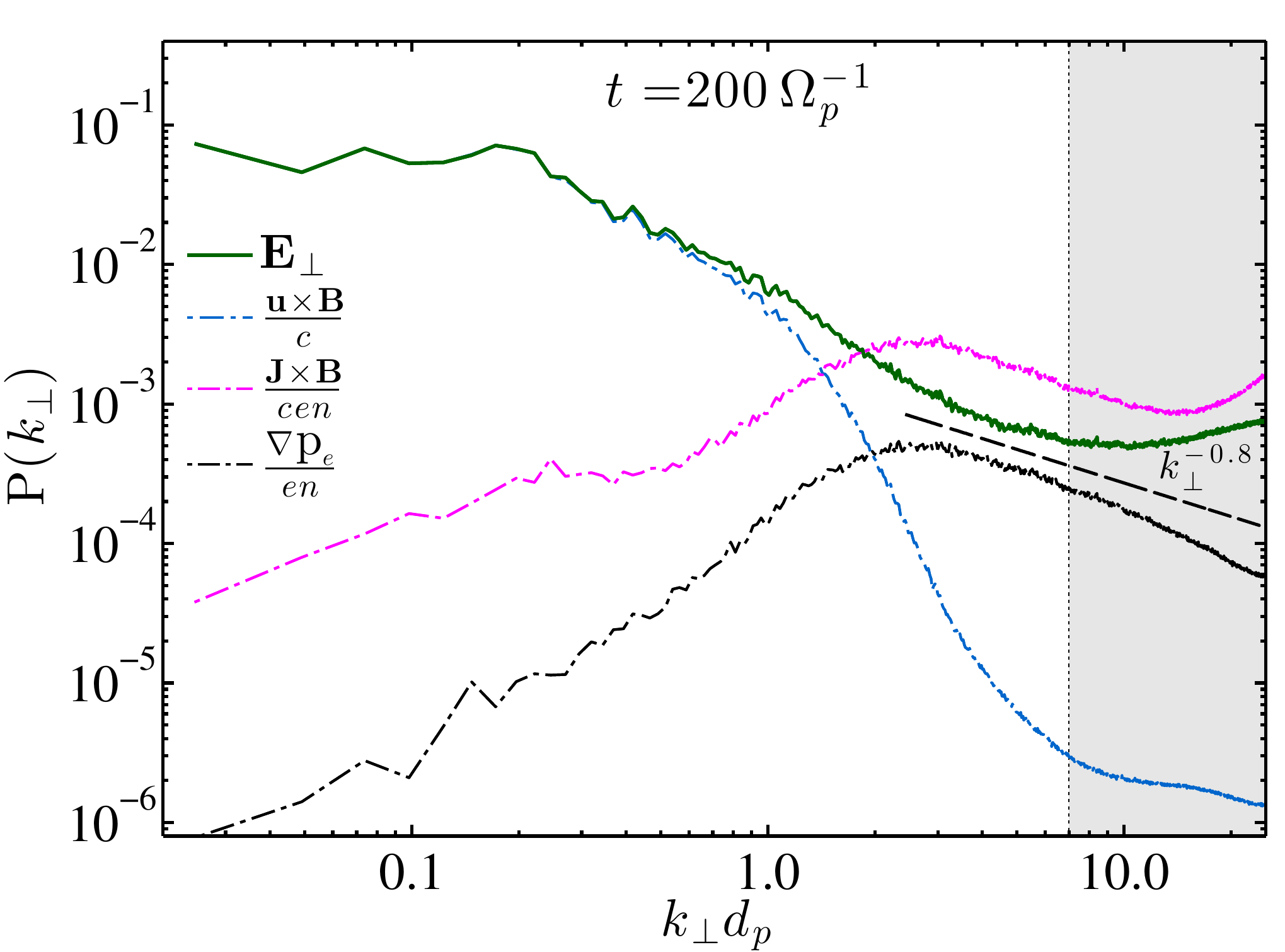}
\caption{Spectrum of the perpendicular electric field, $\vect{E}_\perp$
  (green solid line) and energy associated to the different terms of
  Eq.~\eqref{eq:electricfield} (the term containing the resistive
  coefficient is negligible).  A power law with a spectral index of $-0.8$
  is also drawn with a dashed black line, as a reference.  The shaded
  grey region marks the range where numerical effects strongly affect
  the shape of $P_{\vect{E}_\perp}$.}
\label{fig:efield}
\end{figure}
\begin{figure}[t!]
\centering
\includegraphics[width=0.48\textwidth]{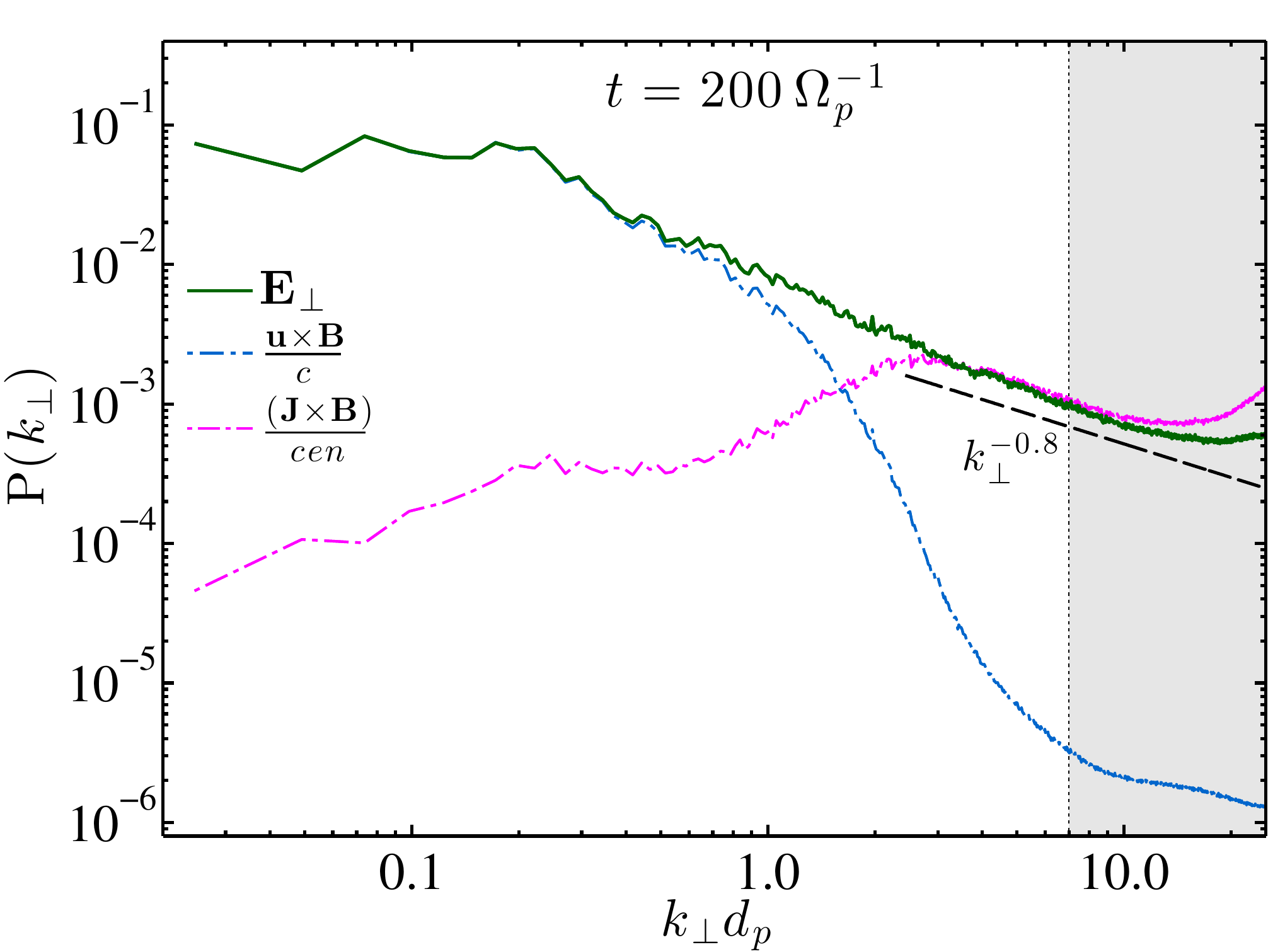}
\caption{Spectrum of the perpendicular electric field, $\vect{E}_\perp$,
 as in Fig.~\ref{fig:efield}, for the case with $\beta_e=0$ ($\bnabla
\mathrm{p}_e=0$).}
\label{fig:efield_beta0}
\end{figure}

Finally, the spectrum of the perpendicular electric fluctuations is
reported in Fig.~\ref{fig:efield}, as a green line.  We choose to pay
particular attention to the electric field for three main
reasons. Uppermost, it is expected to exhibit the most complex
spectrum, since it contains the contributions of four terms having
different relative importance in different ranges of scales. Secondly,
it is the quantity that is mostly affected by both numerical effects
and particle properties, so its behavior needs to be analyzed
carefully, especially at small scales. Lastly, no consistent
observational data about the properties of $\vect{E}$ are available
yet, so making predictions about the shape of its spectrum can be 
relevant for future analysis.  We recall here that, starting
from the Vlasov-fluid equations and assuming that the electrons act as
a massless, charge-neutralizing fluid, the electric field can be
computed from the generalized Ohm's law as
\begin{equation}
  \vect{E} = \underbrace{ - \frac{ \vect{u} \times \vect{B} }{c} }_{\displaystyle
    \vect{E}_\mathrm{MHD}} + \underbrace{ \frac{ \vect{J} \times \vect{B} }
    {cen}}_{\displaystyle \vect{E}_\mathrm{Hall}} \underbrace{-
    \frac{ \bnabla \mathrm{p}_e }{en}}_{\displaystyle \vect{E}_{\mathrm{p}e}} +
  \underbrace{\eta \, \vect{J}}_{\displaystyle\vect{E}_\eta}.
\label{eq:electricfield}
\end{equation}
In Fig.~\ref{fig:efield}, together with $P_{\vect{E}_\perp}$ obtained
from numerical data, we also report the energy associated to the first
three terms of Eq.~\eqref{eq:electricfield}, computed a-posteriori and
drawn with cyan, magenta and black dot-dashed lines respectively (the
contribution from the resistive term is negligible at all scales,
since the resistive coefficient $\eta$ is $5 \times 10^{-4}$.)  At
large scales, $P_{\vect{E}_\perp}$ is clearly dominated by the MHD term,
$\vect{E}_\mathrm{MHD}$, which is essentially perpendicular to
$\vect{B}_0$, since its leading contribution comes from $\vect{u}_\perp
\times \vect{B}_0$ ($\vect{u}_\perp \times \vect{B}_\parallel$ and
$\vect{u}_\parallel \times \vect{B}_\perp$ are both of the second-order in
the fluctuations). Therefore, $P_{\vect{E}_\perp}$ follows strictly
the spectrum of the perpendicular velocity fluctuations
(cf. Fig.~\ref{fig:spectra}).
Approximately at $k_\perp d_p \sim 0.5$, these two spectra decouple,
since the second and third terms of Eq.~\eqref{eq:electricfield} start
contributing.  

We can accurately analyze the Hall term, $\vect{E}_\mathrm{Hall}$, by
considering its perpendicular and parallel components separately.  The
former is of the first-order in the fluctuations, being led by
$(\vect{k_\perp} \times \vect{B}_\parallel) \times \vect{B}_0$ (other
contributions are quadratic in $\vect{B}_\parallel$ and
$\vect{B}_\perp$, and therefore negligible). On the contrary, the
latter is only of the second-order in the fluctuations, having the only
contribution from $(\vect{k_\perp} \times \vect{B}_\parallel) \times
\vect{B}_\perp$. Therefore, we expect the Hall term to be negligible
at large scales, where $\vect{J}_\perp$ is small, and to exhibit a
power-law behavior at small scales, with spectral index $\sim -0.8$
following from $E_{\textrm{Hall}} \propto k^2_\perp
P_{\vect{B}_\parallel}$. Indeed, this is what we observe in
Fig.~\ref{fig:efield} (compare the magenta dot-dashed line with the
reference dashed black line).  The electron pressure gradient term,
$\vect{E}_{\mathrm{p}e}$, has only perpendicular components by
construction (our 2D computational domain is perpendicular to
$\vect{B}_0$). In the inertial range, it is of course negligible, the
spectrum of the density fluctuations being essentially flat (compare with
Fig.~\ref{fig:spectra}).  On the contrary, at small scales, it is
expected to give a contribution $P_{\bnabla n} \propto k^2_\perp
P_{n}$, which has exactly the same slope as the contribution from the
Hall term, since the spectra of the density fluctuations and of the parallel
magnetic fluctuations have the same spectral index of $-2.8$ at
sub-proton scales. This is indeed what we observe in
Fig.~\ref{fig:efield}, where the contribution from the electron 
pressure gradient term is drawn with a black dot-dashed line.  We would expect
a similar behavior for $P_{\vect{E}_\perp}$ at sub-proton scales,
i.e., a power law with a spectral index of $\sim -0.8$, and we observe a
hint of a similar scaling in the range $2 \lesssim k_\perp d_p \lesssim 7$.

The spectrum of the electric field fluctuations is the
most affected by numerical effects among all the considered spectra,
since the computation of $\vect{E}$ involves both other fields
($\vect{u}$ and $\vect{B}$) and derivatives ($\bnabla \times \vect{B}$
and $\bnabla n$), as shown by Eq.~\eqref{eq:electricfield}.  We
already noticed that $P_{\vect{B}}$ suffers from an accumulation of
energy at small scales, which is only a numerical artifact, and so does
$P_{\vect{J}}$ (cf. Fig.~\ref{fig:spectra2}). Moreover, derivatives in
the numerical code are computed as finite differences, thus they are
not able to recover very precise quantities at very small scales. For
all these reasons, we cannot extract any robust information about the
spectrum of the electric field at high wavevectors. In order to
emphasize this, we choose to mark the ``non-safety area'', which we
estimate as $k_\perp d_p \gtrsim 7$, with a shaded gray region in
Fig.~\ref{fig:efield}.

Under particular conditions, i.e., $T_e=0$, one can obtain a better
defined scaling for the electric field. In this case, the electron
pressure gradient term, $\vect{E}_{\mathrm{p}e}$, in the Ohm's law
is zero (see Eq.~\eqref{eq:electricfield}). Consequently, the level of the
electric field spectrum at small scales is higher, since it is 
supported only by the Hall term $\vect{E}_\mathrm{Hall}$. This can be seen in
Fig.~\ref{fig:efield_beta0}, where we show the same analysis of the
electric field spectrum as in Fig.~\ref{fig:efield}, but for a case
with $\beta_e=0$ and all the other parameters set as in Run A.  The
electric field spectrum is now the sum of only two main contributions,
$\vect{E}_\mathrm{MHD}$ and $\vect{E}_\mathrm{Hall}$. No qualitative
changes are introduced with respect to the case with a finite electron
temperature; the former term dominates the spectrum at MHD scales,
while the latter is responsible for the flattening of the electric
field at ion scales. The important difference with respect to Run A, is
that $P_{\vect{E}_\perp}$ displays now a well defined power-law slope
with an index of $-0.8$, consistent with the expectation. We expect
the same slope also for the finite $T_e$ case of Run A, in the absence
of the numerical limitations discussed above.

\section{Role of numerical parameters}
\label{sec:parameters}
\subsection{Spectral properties}
\label{subsec:stability}
\begin{figure}
\centering
\includegraphics[width=0.47\textwidth]{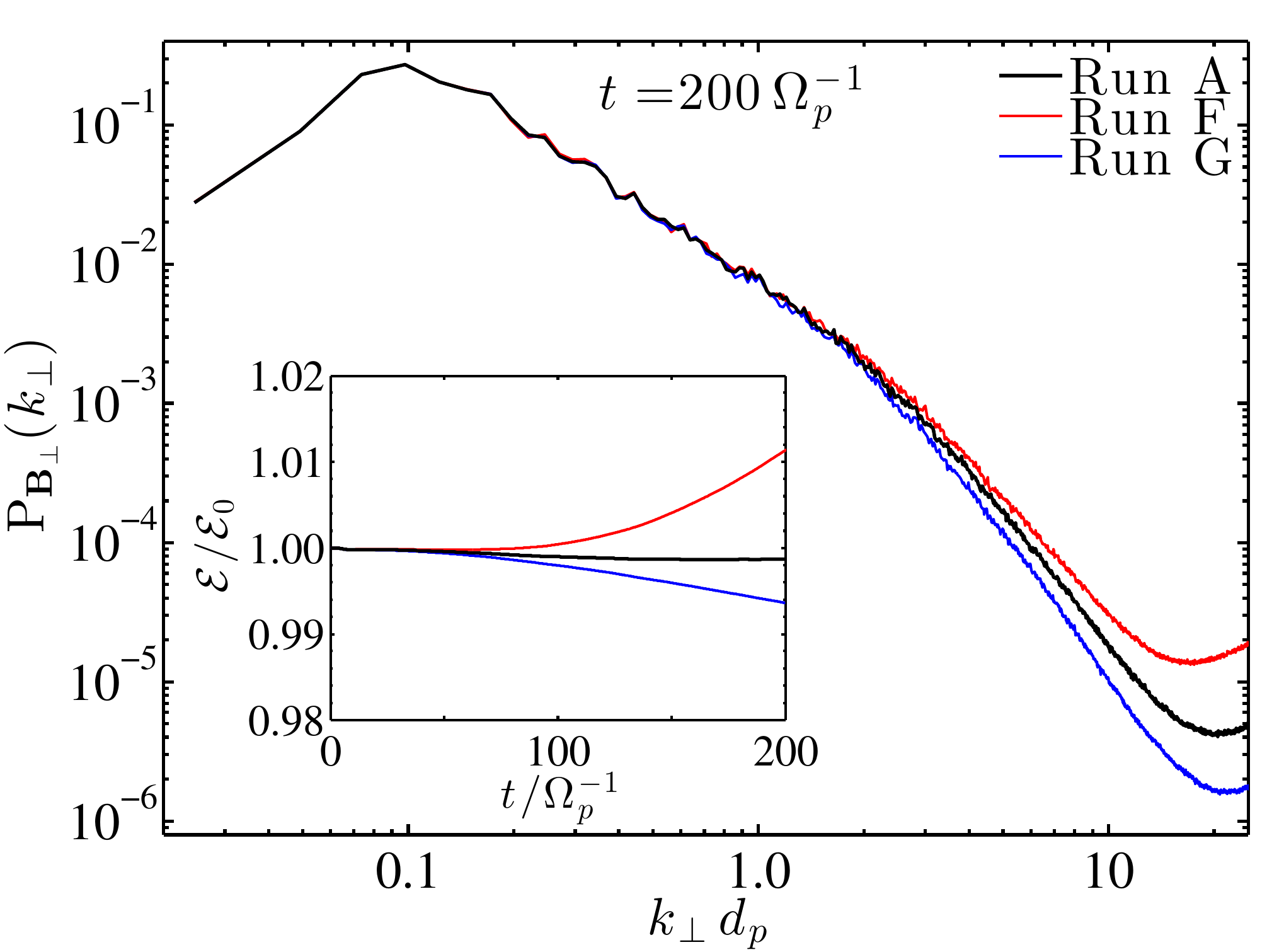}\\ 
\caption{Power spectra of the perpendicular magnetic  fluctuations,
  $\vect{B}_{\perp}$, for simulations with different values of
  the resistivity, i.e., Run A, ($\eta = 5 \times 10^{-4}$, black
  line), Run F ($\eta = 1 \times 10^{-4}$, red line) and Run G ($\eta
  = 1 \times 10^{-3}$, blue line).}
\label{fig:compare_resistivity}
\end{figure}
As mentioned, a small numerical resistivity, $\eta$, has been implemented
in all runs (see Table \ref{tab:modelslist}). A proper level of
resistivity is mandatory in order to prevent an accumulation of the
energy in magnetic fluctuations at small scales, due to numerical
errors.  Runs F and G have been used to fine-tune the resistivity
coefficient, starting from an order-of-magnitude estimate, and
are characterized by the values $\eta = 1 \times 10^{-4}$ and $1 \times
10^{-3}$, respectively.
In Fig.~\ref{fig:compare_resistivity}, we show the corresponding
spectra of the perpendicular magnetic fluctuations, $\vect{B}_\perp$, at
$t = 200 \, \Omega_p^{-1}$, in comparison with Run A 
($\eta = 5 \times 10^{-4}$).  For the setting adopted, the
dissipative scale for the under-resistive case (Run F) can be
estimated as $k_{\textrm{dis}} d_p \sim 35$, i.e., smaller than the
scale corresponding to the employed resolution. As a consequence, this
value of the resistivity is not high enough to remove the energy
excess at small-scales, as also demonstrated by the shape of the
spectrum of Run F in the sub-proton range.
The over-resistive simulation (Run G) corresponds to the opposite case,
where $k_{\textrm{dis}} \, d_p \sim 6$, thus well inside the range of
wavevectors resolved in the simulation, making $P_{\vect{B}_\perp}$
decreasing exponentially below the ion-break, a clear indication of a
too strong dissipative damping at sub-proton scales.  Lastly, the
intermediate case, i.e., Run A, is expected to introduce a
dissipation scale $k_{\textrm{dis}} \, d_p \sim 10$, allowing 
for the best description of the spectrum of $\vect{B}_\perp$. Indeed,
this is observed to follow a power-law scaling with a spectral index of $-3$
for roughly a decade after the break (cf. Fig.~\ref{fig:spectra}), 
in good agreement with solar wind spectra from observational data.  
Therefore, we decided to adopt $5 \times 10^{-4}$ as the optimal
value for the resistive coefficient $\eta$.

As a further confirmation for the adequacy of our choice for $\eta$,
in the insert of Fig.~\ref{fig:compare_resistivity} we also show the
time evolution of the total energy ${\cal E}$, normalized
to its initial value, for the same three simulations.  In all three
cases, the total energy stays constant for $ t \lesssim 70 \,
\Omega_p^{-1}$, while a different behavior is observed at later times.
When the resistivity is too low (RunF, red line) ${\cal E}$ 
grows significantly, due to the inefficient control
of energy at small scales, and such an increase is already of the
order of $\sim 1\%$ at $t = 200 \, \Omega_p^{-1}$.  On the contrary,
when $\eta$ is too high (Run G, blue line) the action of resistivity
is too strong.  Note that part of the energy subtracted from the
magnetic fluctuations would also go into electron heating but, since
the hybrid approximation does not provide an evolution for the
electron temperature, such energy is not taken into account, and is
then lost by the system, resulting in a net decrease of ${\cal E}$. 
The value $\eta = 5 \times 10^{-4}$ is the one which
better ensures the conservation of the total energy, with a relative
difference between the beginning and the end of the simulation, i.e., $t =
500 \, \Omega_p^{-1}$, of about 0.3\%.  Also note that, although the
shape of $P_{\vect{B}_\perp}$ at sub-proton scales is quite strongly
affected by the resistivity, the power-law scaling in the inertial
range and the position of the spectral break are not, assuring the
reliability of the spectra shown in Fig.~\ref{fig:spectra} and
Fig.~\ref{fig:spectra_comp}.

\begin{figure}
\centering
\includegraphics[width=0.47\textwidth]{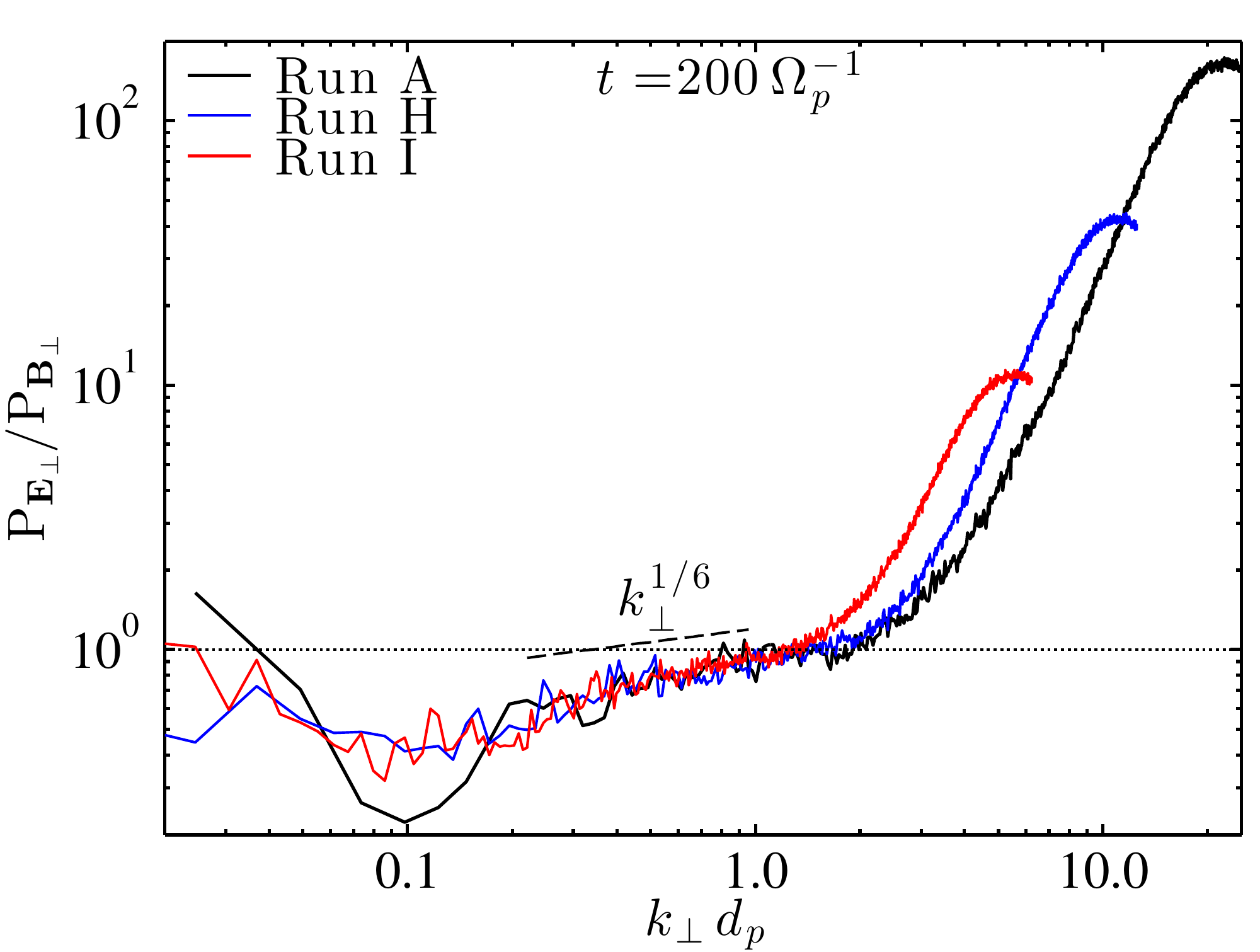}\\
\caption{Ratio between the omnidirectional spectra of perpendicular
  electric fluctuations, $P_{\vect{E}_{\perp}}$, and perpendicular
  magnetic fluctuations, $P_{\vect{B}_{\perp}}$, for simulations with
  different resolution: Run A, with $\Delta x =
  0.125 \, d_p$ (black line), Run H, with $\Delta x = 0.25 \, d_p$ (blue line), 
  and Run I, with $\Delta x = 0.5 \, d_p$ (red line).}
\label{fig:compare_resolution}
\end{figure}
We also investigated the stability of omnidirectional spectra versus
the spatial resolution. This was done by varying $\Delta x$, while
keeping fixed the number of grid points (the total box length is then
larger for larger grid spacing), as well as the amplitude of the
initial magnetic fluctuations, $B^{\textrm{rms}}$.  Run H and I
implement $\Delta x = 0.250$ and $0.500 \,d_p$, respectively (see Table
\ref{tab:modelslist}).  For both these runs, the value of the
resistivity coefficient was suitably rescaled in order to get
dissipation at the proper scales (note that $\eta \propto \Delta x$
under these conditions).

In Fig.~\ref{fig:compare_resolution}, the ratio between the
omnidirectional spectra of the perpendicular electric and magnetic
fluctuations, $P_{\vect{E}_{\perp}}/ P_{\vect{B}_{\perp}}$, is
compared between Run A, H and I.  For all the three runs, this ratio
exhibits the same scaling in the inertial range, following a power law
with a spectral index of $1/6$.  This is a direct consequence of the
different scaling for the magnetic field ($-5/3$) and the velocity
($-3/2$) in the ideal MHD regime -- where also $P_{\vect{E}_{\perp}}\sim
P_{\vect{u}_{\perp}}$ -- leading then to a spectral index of
$-3/2+5/3=1/6$ for their ratio.  As the other terms in the generalized
Ohm's law became important at ion scales, the
$P_{\vect{E}_{\perp}}/P_{\vect{B}_{\perp}}$ ratio increases
significantly at smaller $k$s.  Interestingly, increasing $\Delta x$
from $0.125$ (blue) to $0.250 \,d_p$ (black) does not produce a change
in the scale at which $P_{\vect{E}_{\perp}}$ exceeds
$P_{\vect{B}_{\perp}}$, and this break is observed to occur at
$k_\perp \, d_p \sim 2$ in both cases. Moreover, the two curves
exhibit similar slopes in the sub-proton ranges.  This is a
confirmation that the estimate of $\eta$ for the two simulations was
correct and that the raise in
$P_{\vect{E}_{\perp}}/P_{\vect{B}_{\perp}}$ is physical and well
captured by the runs.  On the other hand, when employing a lower
resolution, $\Delta x = 0.500$ (red), the break seems to occur at
slightly larger scales. This is likely a consequence of the reduction
of the resolution at small scales: in Run I, the break and the
dissipative scale are not well separated in Fourier space, so that
subtracting energy at the smallest scales via resistive dissipation
also affects the shape of the spectra around $k_\perp \, d_p \sim 2$,
where the break occurs.  This is an evidence that the scaling for the
spectra discussed and shown in Fig.~\ref{fig:spectra}, continue to
hold at lower spatial resolution, but also that $\Delta x \gtrsim
0.500 \,d_p$ is not sufficient to properly explore the physical
behavior at sub-ion scales.
\begin{figure}[t!]
\centering
\includegraphics[width=0.48\textwidth]{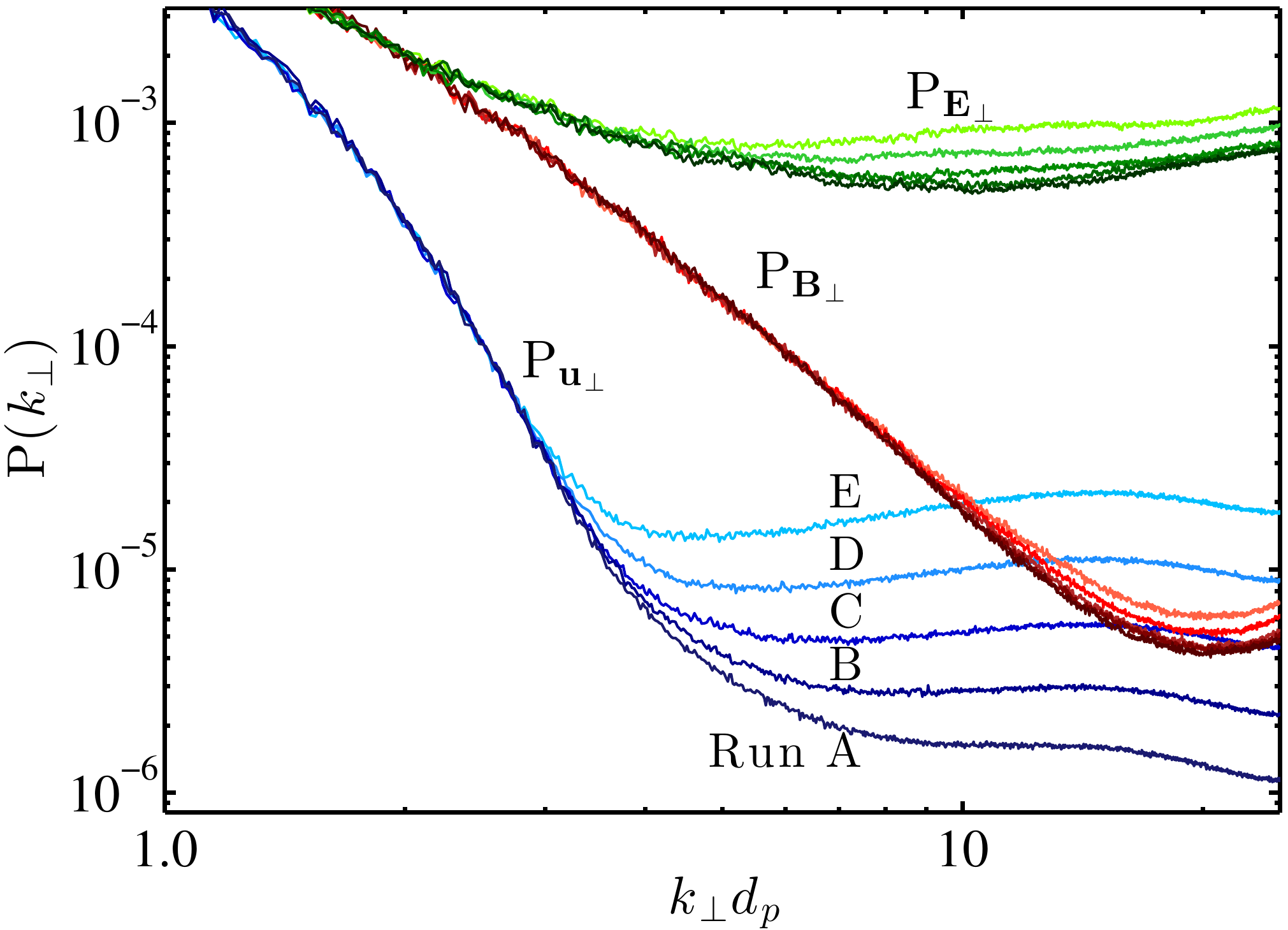}
\caption{Zoom of the power spectra of the perpendicular magnetic,
  electric and proton bulk velocity fluctuations (drawn with red,
  green and blue lines, respectively) at small scales, where the
  contribution of numerical noise is not negligible. Lines with
  different shades of the same color correspond to simulations with
  different amounts of ppc, ranging from 500 to 8000, with darker
  colors being associated with a higher number of particles.}
\label{fig:spectranoise}
\end{figure}

Finally, the importance of employing a high number of particles was
investigated, by keeping all the same parameters as for Run A
except for varying the number of ppc from 4000 to 500
in steps of a factor of 2 (Runs from B to E in
Table~\ref{tab:modelslist}). In Fig.~\ref{fig:spectranoise}, the spectra
of the perpendicular velocity, magnetic and electric fluctuations are
reported with lines in different shades of blue, red and green
respectively, corresponding to simulations with different numbers of
ppc ranging from 500 (lighter color, Run E) to 8000 (darker color,
Run A).  Increasing the number of pcc from 500 ($\sim 2.0 \times
10^{9}$ total particles) to 8000 ($\sim 3.3 \times 10^{10}$ total
particles) results in an decrease of the noise at small scales of more
than one order of magnitude for the spectrum of the perpendicular velocity fluctuations.
On the other hand, the trend of the spectrum up to the proton inertial length
and slightly above is not affected, and the curve corresponding to our
most accurate simulation (Run A) overlaps with the one with 4000 ppc (Run
B) up to $k_\perp d_p \sim 4$. On the contrary, the spectrum of the
perpendicular magnetic fluctuations is only barely affected by the
numerical noise when the number of particles is sufficiently high,
since the curves for 4000 and 8000 ppc are almost indistinguishable,
proving that the number of ppc employed in Run A is sufficient to
get reliable results for $P_{\vect{B}_\perp}$ up to $k_\perp d_p \sim 10$.  
Lastly, the spectrum of the perpendicular electric fluctuations shows
a dependence on the number of particles at scales $k_\perp d_p \gtrsim
4$, but the curves for 4000 and 8000 ppc are quite close to each other
even at smaller scales. However, as mentioned, $P_{\vect{E}_\perp}$ is
influenced by different sources of numerical noise, and all
contribute in affecting the spectrum at small scales.  We would like
to stress that the evaluation of the noise due to the finite number of
ppc only represents a lower limit of the overall noise, and therefore
our previous choice of marking a shaded grey area for $k_\perp d_p
\gtrsim 7$ in Fig.~\ref{fig:efield} is not in contrast with these
results.

\subsection{Proton heating}
\label{subsec:heating}
Fig. \ref{fig:time_evolutions} shows that some particle heating is
observed during the turbulent activity. Some care must be used in the
interpretations of this result, since it may be significantly affected
by some of the numerical settings. Therefore, we carefully consider
the properties of the proton heating in this subsection.  

\begin{figure}[t!]
\centering
\includegraphics[width=0.48\textwidth]{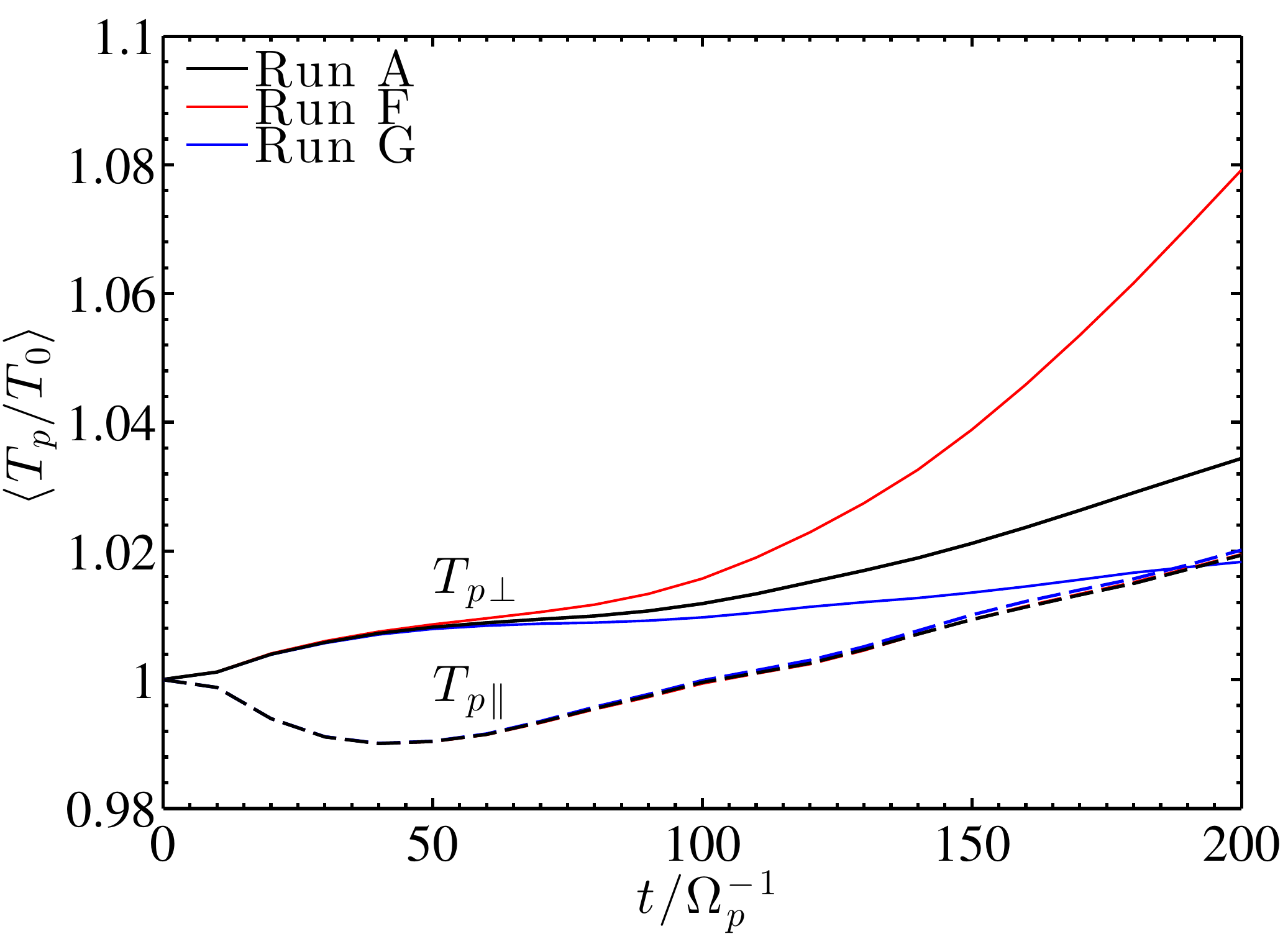}
\caption{Time evolution of the parallel and perpendicular proton
  temperature, $T_{p\parallel}$ and $T_{p\perp}$, respectively, normalized to
  the initial common value, $T_0$. The evolution is here shown for different
  values of the resistive coefficient (see Table~\ref{tab:modelslist}).
  Solid and dashed lines are used for $T_{p\perp}$ and $T_{p\parallel}$,
  respectively.}
\label{fig:Tperp_eta}
\end{figure}

Resistivity is observed to play a fundamental role in determining the
overall proton heating, $\Delta T_{\|,\perp}=\langle
T_{\|,\perp}\rangle -T_0$, and the proton temperature anisotropy, $A_p$.
In Fig.~\ref{fig:Tperp_eta}, we show the time evolution of the
perpendicular $T_{p\perp}$ and the parallel $T_{p\parallel}$ 
proton temperature, in solid and dashed lines, respectively,
corresponding to different values of the resistive coefficient $\eta$
(Run A, F and G of Table~\ref{tab:modelslist}).  The time evolution
of $T_{p\parallel}$ is observed to be not affected almost at all by the
resistivity, showing an early decrease up to $t \sim 50 \,
\Omega_p^{-1}$ and then an increase with an almost constant rate, as
was already shown for Run A in Fig.~\ref{fig:time_evolutions}.  The
situation is different for $T_{p\perp}$, since its behavior for
different values of $\eta$ is the same only during the initial readjustment
of the system, while it starts to differ after $t\sim 40
\Omega_p^{-1}$.  At later times, $T_{p\perp}$ exhibits a growth rate
very similar to that of $T_{p\parallel}$ for Run A (black), so no
preferential perpendicular or parallel heating is observed.  In
particular, $\Delta T_{p\perp} /T_0$ at $t = 200 \, \Omega_p^{-1}$ is
about 3.5\%, while the corresponding $\Delta T_{p\parallel} / T_0$ is
about 2\%.  When $\eta$ is lower (Run F, red), $T_{p\perp}$ grows with
a much faster rate than $T_{p\parallel}$ for $t \gtrsim 50 \,
\Omega_p^{-1}$, generating a strong preferential heating in the
perpendicular direction, $T_{p\perp}$ being about 8\% greater than the
initial value at $t = 200 \, \Omega_p^{-1}$.  On the contrary, when
$\eta$ is higher (Run G, black lines), $T_{p\perp}$ grows much slower,
being overcome by $T_{p\parallel}$ just before $t = 200 \,
\Omega_p^{-1}$, leading to $A_p < 1$ at later times.  The amount of
perpendicular heating observed is then significantly related to the
presence of an excess of fluctuations at small scales, and can be then
therefore largely overestimated, or underestimated, if an incorrect
value of the resistivity is adopted.
\begin{figure}[t!]
\centering
\includegraphics[width=0.48\textwidth]{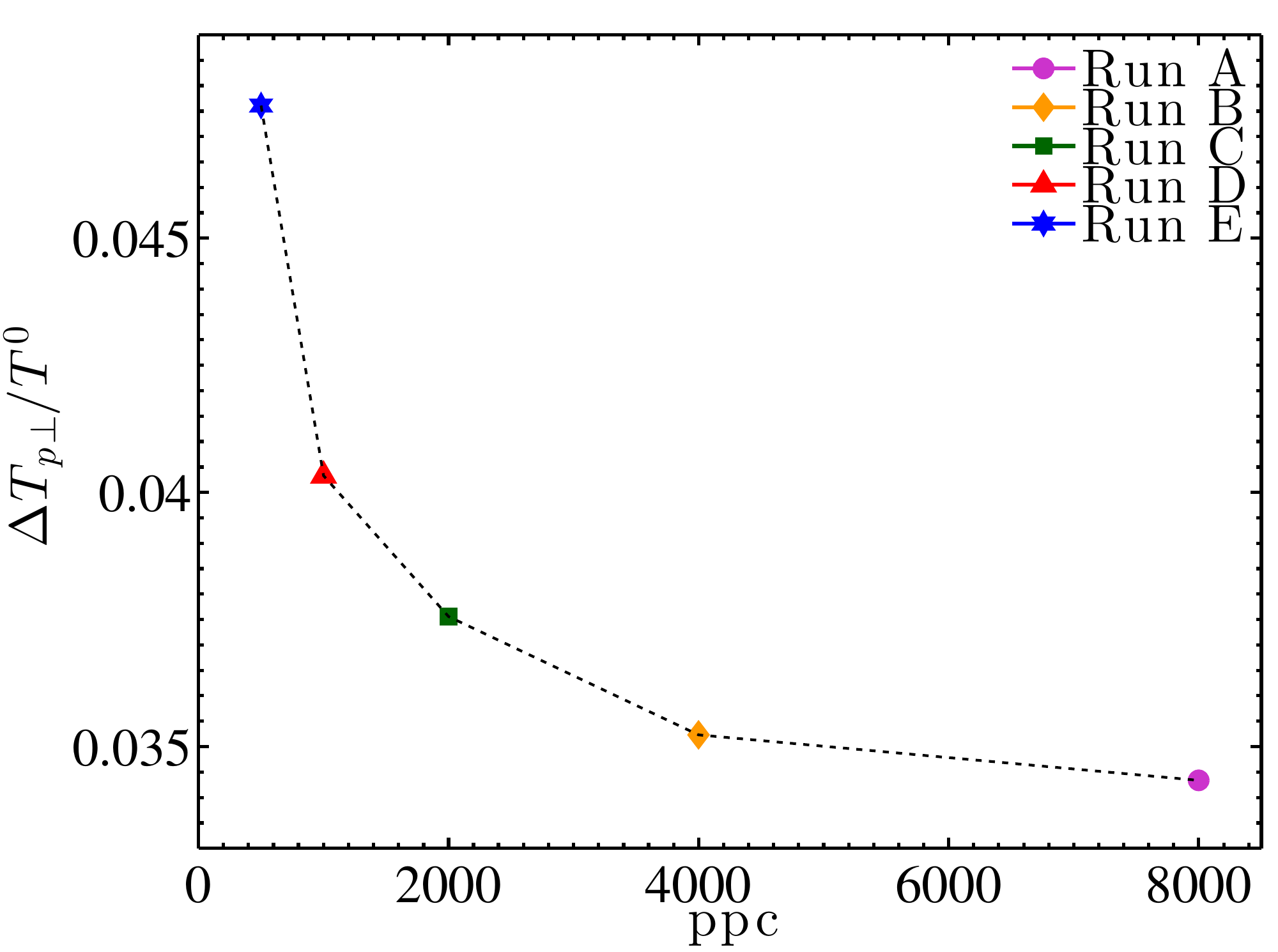}
\caption{Perpendicular proton heating, $\Delta T_{p\perp} / T_0$, computed
  at $t = 200 \, \Omega_p^{-1}$, versus the number of ppc employed,
  ranging from 8000 (Run A) to 500 (Run E).}
\label{fig:PerpHeatingPPC}
\end{figure}

The perpendicular proton heating is also found to be strongly affected
by the number of particles employed.  In Fig.~\ref{fig:PerpHeatingPPC},
we report the ratio $\Delta T_{p\perp} / T_0$, where $ \Delta
T_{p\perp} = \langle T_{p\perp}\rangle - T_{0}$ is computed at $t =
200 \, \Omega_p^{-1}$, versus the number of ppc for Runs from A to E
(see Table~\ref{tab:modelslist}).
At this time, $\Delta T_{p\perp}/T_0$ is clearly
higher when employing a lower number of ppc. In particular, while a
good convergence towards a value of $\sim 3.4\%$ is observed when
increasing the number of ppc from 4000 to 8000, this quantity clearly
diverges when few particles are employed, reaching 4.8\% for 500 ppc.
Moreover, the difference in $T_{p\perp}$ between Run A and Run E tends
to further increase at later times.  The main result of this analysis
is that the use of a high number of ppc is mandatory when trying to give
an estimate of $T_{p\perp}$, which could be largely overestimated
otherwise. On the contrary, the parallel proton temperature is found
to be largely independent from the number of particles (the relative
difference between Run A and Run E is lower than 0.1\% at $t = 200 \,
\Omega_p^{-1}$).

To conclude our analysis, we find that the spatial resolution $\Delta
x$ does not seem to affect significantly the proton heating, provided
that the value of the resistivity is suitably set as discussed in
Subsection~\ref{subsec:stability}.  Differences between $T_{p\perp}$
at $t = 200 \, \Omega_p^{-1}$ for Runs A, H, and I are less than
0.6\%.  The dependency of $T_{p\parallel}$ on the spatial resolution
is also negligible.

\section{Summary and conclusions}
\label{sec:conclusions}
In this work, we have presented properties of turbulence in a
magnetized collisionless plasma by means of two-dimensional hybrid
PIC simulations, extending the results of \citetalias{Franci_al_2015a}.  
Remarkably, our simulations implement
a high number of collocation points ($2048 \times 2048$) and a very
high number of particles (up to $8000$ ppc), covering a large
simulation domain (not less than $L_{\textrm{box}} = 256 \, d_p$)
with a fine spatial resolution. This enables to self-consistently
describe the evolution of turbulence over three orders of magnitude in
wavevectors, and to fully capture its transition from fluid-like MHD
scales to kinetic sub-ion scales, by using a single simulation (see 
\citetalias{Franci_al_2015a}).

The adopted initial conditions consist of balanced and equipartitioned
magnetic and velocity fluctuations, i.e., with zero cross helicity and
zero residual energy.  The onset of a turbulent cascade appears quite early
during the simulations ($t \sim 200 \, \Omega_p^{-1}$, corresponding
to approximatively $10$ nonlinear times $t_{\textrm{NL}}$), i.~e.,
when most of the initial modes have started to partake into the
cascade. In physical space, the activity of turbulence is characterized
by magnetic field coherent structures, vortices, and strong and
localized current sheets at smaller scales.

Generation of coherent structures associated to intermittency is
observed as turbulence evolves through MHD to sub-proton scales. PDFs
of increments of a perpendicular component of the magnetic fluctuations at
$t = 200 \, \Omega_p^{-1}$ exhibit a deviation from the normal
distribution at all scales. This is small in the inertial range,
becoming larger in correspondence of the spectral break at $k_\perp \,
d_p \sim 2$, while at $k_\perp \, d_p \sim 2$ the PDF has a much leaner
shape with long non-zero tails.
The corresponding excess kurtosis confirms this behavior. It is
very small around the injection scale, since part of the MHD range
fluctuations still acts as an energy reservoir for turbulence at $t =
200 \, \Omega_p^{-1}$, while it increases through the inertial
range. Moreover, we observe a further increase at smaller scales.
Observational data give no firm results about the behavior of this
quantity at different scales
\citep{Alexandrova_al_2008b,Kiyani_al_2009,Kiyani_al_2013,Wu_al_2013,Chen_al_2014a}.
Nevertheless, when shown, all previous simulations observe an increase
of the kourtosis at smaller scales
\citep[e.g.,][]{Dmitruk_Matthaeus_2006a,Wan_al_2012,Karimabadi_al_2013,Wu_al_2013}.

When looking at the spectra of the relevant quantities, two clear
distinct turbulent regimes are observed. At larger scales, the magnetic
field follows a Kolmogorov $-5/3$ power law, while the velocity has a
spectral index of $-3/2$, which is characteristic of a Iroshnikov-Kraichnan turbulence. 
An excess of magnetic energy with respect to the kinetic energy is observed
throughout the inertial range. The two different scalings for the magnetic
and velocity fluctuations, often observed in the solar wind
\citep{Podesta_al_2006a,Podesta_al_2007,Tessein_al_2009,Salem_al_2009,Chen_al_2011a},
are very stable in time. They appear at the maximum of the turbulent
activity and persist throughout all the simulations, as the energy
reservoir at large scales is able to sustain and maintain the
cascade. In \citetalias{Franci_al_2015a}, we showed that such magnetic and velocity scaling
are also combined with a spectral index of $-2$ for the residual energy,
in agreement with observations in the solar wind
\citep{Chen_al_2013b}. Incompressible MHD \citep{Muller_Grappin_2005}
and Reduced MHD \citep{Boldyrev_al_2011} only partially reproduce such
scaling. In our simulations, the 2D geometry and the presence of compressibility
may play a role in setting the different scaling.

A clear transition in the spectra is observed at scales 
$k_\perp d_p \gtrsim 1$, with a change in the spectral indices of all fields.
In particular, the spectrum of the perpendicular magnetic fluctuations
steepens at $k_{\perp} \, d_p \sim 2$, following a power law with
spectral index $\sim -3$ for another decade.  The location of the
break does not show any significant dependence on the number of
particles, the spatial resolution and the resistivity adopted,
provided that a sufficient number of grid points allows to cover
approximatively a decade at sub-proton scales, i.e., that the scale at
which resistive dissipation acts is sufficiently separated from the
region of the break.  The parallel component of the magnetic field,
together with the density, follows a similar but slightly shallower
slope with a spectral index of $\sim -2.8$, in very good agreement with observations
\citep{Chen_al_2012,Chen_al_2013} and other simulations
\citep{Howes_al_2011,Passot_al_2014}.  As a result, magnetic
fluctuations tend to become isotropic at small scales, resulting in an
increase of the magnetic compressibility, as observed in the solar wind
\citep{Salem_al_2012,Podesta_TenBarge_2012,Kiyani_al_2013}. The
spectrum of the perpendicular velocity fluctuations quickly drops above
$k_{\perp} \,d_p \sim 1$, without any clear power-law trend. The
observation of a spectral index of $-2.8$ has been ascribed to the effect
of the electron Landau damping by previous studies
\citep{Howes_al_2011,Passot_al_2014}, however, this can not be the
case in our simulations, where the electron kinetics is not taken into
account. Alternatively, the presence of coherent structures, such as
current sheets, can produce a steepening of the energy spectra
\citep[e.g.,][]{Wan_al_2012,Karimabadi_al_2013}. The increase of
intermittency at small scales, observed in our simulations, seems
to confirm this path towards the dissipation.  We have to note, however,
that a $-8/3$ power law for the magnetic energy and the density spectra (not
far from the $2.8$ found here) has been also interpreted as related to
the dimensionality (1D or 2D) of the magnetic and the density intermittent
structures, without invoking dissipation \citep{Boldyrev_Perez_2012,Meyrand_Galtier_2013}.

The spectrum of the electric fluctuations is highly dominated by its
perpendicular component. It is strongly coupled to the spectrum of the
perpendicular velocity fluctuations at fluid scales, then it decouples
and flattens, exceeding the spectrum of the perpendicular magnetic
fluctuations and becoming dominant for $k_{\perp} \, d_p \gtrsim 2$.
At large scales, the only contribution comes from the MHD term in
Eq.~\eqref{eq:electricfield}, whose leading term is
\begin{equation}
\vect{E}_\perp \propto \vect{u} \times \vect{B} \sim \vect{u}_\perp 
\times \vect{B}_0.
\end{equation}\label{eq:e_perp}
This corresponds to a power-law scaling
\begin{equation}
P_\vect{E} \propto P_\vect{u_\perp} \propto k_\perp^{-3/2},
\end{equation}
which is observed in the simulations, and is also consistent with
observations \citep{Chen_al_2011b}.  In our case, the main
contributions at sub-proton scales come from the Hall and the
electron pressure gradient terms, since the spectrum of the velocity
fluctuations is observed to drop exponentially at short wavelengths.
The leading terms at these scales are, then,
\begin{equation}
  \vect{J} \times \vect{B} - \bnabla \mathrm{p}_e \propto \bnabla (T_e
  n + \vect{B}_0 \cdot \vect{B}_\parallel )\,.
\end{equation}
Note that, in general, the sum of the two terms inside parentheses would
not necessarily result in a power law for the electric field. However,
since in our simulations both $n$ and $\vect{B}_\parallel$ are observed to
scale with the same power law -- thanks to the strong coupling between
the plasma and the magnetic compressibility -- then the expected spectral
index for the electric field is:
\begin{equation}
P_\vect{E} \propto k^2_\perp P_{\vect{B}_\parallel,n} \propto
k_\perp^{-0.8}\,.
\end{equation}\label{eq:e_subion}
Although it is not possible to directly test this scaling for the
electric field spectrum in Run A, individual terms in
Eq. \eqref{eq:e_subion} follow well the prediction (see
Fig. \ref{fig:efield}).  Moreover, we were able to show that when
assuming $T_e=0$ (i.e., setting to zero the electron pressure gradient term in
the Ohm's law), then the electric field spectrum -- hence dominated by
the Hall term -- follows a $k_\perp^{-0.8}$ scaling in the sub-proton
range (Fig. \ref{fig:efield_beta0}), being only very slightly affected
by numerical effects at very small scales ($k_\perp \, d_p \gtrsim 10$).

As a result of the interaction of particles with the turbulent
fluctuations and small scale structures, we observe an overall
parallel and perpendicular heating with similar rates, so that the
temperature of the plasma remains globally nearly isotropic.  This
behavior can be achieved only if a high enough number of particles is
employed, and the resistivity is properly set in order to assure an
accurate conservation of the total energy and a clear power-law behavior
for the spectrum of the magnetic fluctuations at all scales.  The parallel
temperature, $T_{p\parallel}$, is found to have a very robust evolution,
being essentially independent of the resistivity, the number of
particles, and the spatial resolution employed.  On the contrary, the
time evolution of $T_{p\perp}$ is strongly determined by both the
resistivity and the number of ppc: if too few particles are employed,
or if the resistivity is too low, the perpendicular heating can be
largely overestimated/unphysical. Conversely, when a too strong
value of the resistivity is implemented, the artificial damping of
fluctuations at ion scales can produce a strong
reduction of the perpendicular heating, thus generating an equally
unphysical preferential parallel heating.  This proves that no firm
conclusions can be drawn about the perpendicular heating by turbulence
in hybrid simulations, unless a careful and empirically fine-tuned
choice of all parameters has been taken.

Note, however, that the fact that we do not observe a global preferential
heating does not imply the absence of signatures of localized
preferential deformations of the particle distribution functions, as
suggested by the bottom right panel of Fig.~\ref{fig:turbulence}, where
strong temperature anisotropies ranging from $0.5$ to $1.8$ are
observed. They seem to be concentrated in regions with stronger coherent
structures, identified by the presence of current sheets and a
significant level of vorticity. These results are in agreement with
previous works based on the Vlasov-hybrid approximation
\citep[e.g.,][]{Servidio_al_2012,Perrone_al_2013,Valentini_al_2014,Servidio_al_2014a}.
As the overall heating is rather weak, slow, and nearly isotropic, we
can infer that the local formation of large proton temperature
anisotropies is likely due to energy exchanges between the parallel
and perpendicular directions, and/or to the spatial transport, rather 
than due to the heating.

Solar wind observations show a certain variability of the spectral
properties. In particular, the position of the break at ion scales and
the shape of the magnetic field spectrum around it seems to
depend on the power of magnetic fluctuations \citep{Bruno_al_2014} and
on the plasma beta \citep{Alexandrova_al_2008b, Chandran_al_2009, Chen_al_2014}.
Investigating such a dependence, by exploring the parameter space of
the level of fluctuations and the plasma beta, will be the subject of
a fortcoming paper.

Three-dimensional simulations would be fundamental to further improve
the present study and overcome its limitations, allowing for a more
realistic description of the turbulent cascade.

\acknowledgments

The authors wish to acknowledge valuable discussions with Olga
Alexandrova, Chris Chen, Giuseppe Consolini, Roland Grappin, Frank
L\"offler, and Marco Velli.  This project has received funding from
the European Union’s Seventh Framework Programme for research,
technological development and demonstration under grant agreement no
284515 (SHOCK). Website: \url{http://project-shock.eu/home/}.  This
research was conducted with high performance computing (HPC) resources
provided by the Louisiana State University (allocations hpc$\_$hyrel14
and hpc$\_$hyrel15) and by CINECA (grant HP10CVCUF1).
P.H. acknowledges GACR grant 15-10057S. L.M. was funded by STFC grant
ST/K001051/1. A.V. acknowledges the Interuniversity Attraction Poles
Programme initiated by the Belgian Science Policy Office (IAP P7/08
CHARM).

\bibliographystyle{apj-eid}

\end{document}